\documentclass[12pt]{iopart}

\usepackage{iopams}  
\usepackage{bbm}
\usepackage{mathrsfs}
\usepackage{amsfonts}
\usepackage[colorlinks=true,linkcolor=blue,urlcolor=blue,citecolor=blue,anchorcolor=blue]{hyperref}
\usepackage{graphicx,epstopdf}
\usepackage{subfigure}
\usepackage{epsfig}
\usepackage{dcolumn}
\usepackage{bm}
\usepackage{color}
\usepackage[sort&compress, numbers]{natbib}
\usepackage{amssymb}
\usepackage{xcolor}
\usepackage{braket}
\usepackage{ulem}
\usepackage{float}

\begin{document}

\title{Fast quantum state transfer and entanglement preparation in strongly coupled bosonic systems}
\author{Yilun Xu$^{1,2}$, Daoquan Zhu$^{1}$, Feng-Xiao Sun$^{1,3,*}$, Qiongyi He$^{1,2,3,4,5}$, Wei Zhang$^{6,2,\dagger}$}
\address{$^1$ State Key Laboratory for Mesoscopic Physics, School of Physics, Frontiers Science Center for Nano-optoelectronics, Peking University, Beijing 100871, China}
\address{$^2$ Beijing Academy of Quantum Information Sciences, Beijing 100193, China}
\address{$^3$ Collaborative Innovation Center of Extreme Optics, Shanxi University, Taiyuan, Shanxi 030006, China}
\address{$^4$ Peking University Yangtze Delta Institute of Optoelectronics, Nantong, Jiangsu 226010, China}
\address{$^5$ Hefei National Laboratory, Hefei, Anhui 230088, China}
\address{$^6$ Department of Physics and Key Laboratory of Quantum State Construction and Manipulation (Ministry of Education), Renmin University of China, Beijing 100872, China}
\ead{$^*$sunfengxiao@pku.edu.cn}
\ead{$^\dagger$wzhangl@ruc.edu.cn}

\begin{abstract}
Continuous U$(1)$ gauge symmetry, which guarantees the conservation of total excitations in linear bosonic systems, will be broken when it comes to the strong-coupling regime where the rotation wave approximation (RWA) fails. Here we develop analytic solutions for multi-mode bosonic systems with XX-type couplings beyond RWA, and propose a novel scheme to implement high-fidelity quantum state transfer (QST) and entanglement preparation (EP) with high speed. The scheme can be realized with designated coupling strength and pulse duration with which the excitation number keeps unchanged regardless of the breakdown of the global U$(1)$ symmetry. In QST tasks, we consider several typical quantum states and demonstrate that this method is robust against thermal noise and imperfections of experimental sequence. In EP tasks, the scheme is successfully implemented for the preparation of Bell states and W-type states, within a shortest preparation time.

\end{abstract} 


\submitto{\NJP}
	
\maketitle
\section{Introduction} 
\label{sec:intro}

In the past several decades, the linear transformation of boson modes has been an intriguing topic in quantum optics and quantum information sciences. For example, it plays a crucial role in the problem of Bose samplings~\cite{doi:10.1126/science.1231692,RN94,RN95,PhysRevLett.123.250503,RN96}, which is of particular interest in the study of near-term platform for photonic quantum computing~\cite{PhysRevLett.119.170501}. The linear bosonic transformation can be effectively realized by optical beam splitters and wave plates in optical systems under the protocol raised by Knill, Laflamme and Milburn~\cite{RN93,RevModPhys.79.135,PhysRevLett.73.58}, or by controllable pulse manipulation in superconductor-waveguide systems~\cite{PhysRevLett.118.133601,RN71,PRXQuantum.2.030321,PhysRevX.7.011035,PhysRevApplied.17.054021,PhysRevLett.130.050801}. Experimental progresses in this direction have facilitated the development of various quantum information tasks, including quantum state transfer (QST), entanglement preparation (EP) and entanglement distribution~\cite{PRXQuantum.2.030321,PhysRevLett.130.050801,PhysRevX.7.011035,PhysRevLett.108.153604,PhysRevApplied.17.054021,PhysRevA.87.012339}.

As an important step of quantum information processing, QST aims to transfer an arbitrary quantum state from the sender side to the receiver side with high fidelity and fast speed. Recently, to embrace the noisy intermediate-scale quantum era for quantum internet frameworks~\cite{Kimble,RN70,RevModPhys.82.1041,PhysRevLett.130.050801,PhysRevLett.78.3221}, much effort has been made to implement QST tasks in various physical systems, including atom-cavity systems~\cite{PhysRevLett.78.3221,Ritter,PhysRevLett.118.133601,PhysRevA.50.R3589}, superconducting circuits~\cite{PhysRevX.7.011035,PhysRevApplied.10.054009,PhysRevLett.118.133601,RN71,PRXQuantum.2.030321,PhysRevApplied.17.054021}, photonic systems~\cite{PhysRevB.84.014510,PhysRevLett.95.170501,Xiao-Ling}, mechanical oscillators~\cite{Brown}, opto-mechanical cavities~\cite{PhysRevLett.108.153604,PhysRevA.92.053804,Zeng:20,PhysRevA.86.021801,Parkins_1999,Barzanjeh2022Optomechanics} and so on~\cite{Kane,PhysRevLett.106.040505,PhysRevA.87.012339,Kandel,PhysRevLett.99.093901}. Meanwhile, the generation of entangled states, as the first and fundamental step to realize quantum algorithms and manifest the so-called quantum supremacy, has been widely studied~\cite{PRXQuantum.2.030321,RN97,RN98,RN100,RN101,PhysRevApplied.17.054021}. By manipulating the bosonic transformation matrix, entanglement resource can be generated and distributed among different ports of user. In the weak-coupling regime where the rotation wave approximation (RWA) can be safely adopted, QST and EP can be accomplished with high fidelity but slow speed~\cite{PhysRevX.7.011035,PRXQuantum.2.030321,PhysRevLett.118.133601,PhysRevApplied.17.054021}. On the other hand, as one tries to increase the processing speed and pushes to the strong-coupling regime, the breakdown of the global U$(1)$ symmetry will lead to the failure of RWA, resulting in the deviation of the desired bosonic transformation. In such a case, one has to go beyond RWA and develop new schemes to suppress errors brought by the broken U$(1)$ symmetry. 

In this paper, we develop a scheme to realize linear bosonic transformation for QST and EP in the strong-coupling regime. By formulating an analytic solution of the widely-used multi-mode model, we find that the total excitation number will be preserved for accurately controlled coupling strength and at certain discrete points of time, in despite of  the broken U$(1)$ symmetry. For QST tasks involving two terminal modes and one intermediate channel mode, we obtain an analytical tradeoff relation between the transfer speed and the fidelity. Taking several typical states as examples, we further demonstrate that our method outperforms the traditional one derived from RWA with lower infidelity and absolute robustness against thermal noise of the intermediate mode. The fidelity can be further improved by applying a simple local operation to compensate the phase rotation induced by the strong-coupling. And we also show the potential to extend our protocol to multi-mode QST tasks through an example of transferring an arbitrary W-type state. For EP tasks, we show that typical entangled states such as Bell states and multi-mode W-type states~\cite{PhysRevA.50.R2799} can be successfully generated using the proposed scheme, where the fastest preparation time with our scheme can also be derived. The degree of entanglement can be modified by changing the coupling strengths of different modes.

\section{Model}
\label{sec:model}

We start from a widely adopted model where $n$ oscillator-encoded qubits are coupled to an intermediate channel mode with an XX-type coupling~\cite{Brown,Ockeloen-Korppi,PhysRevA.84.052327,PhysRevA.92.053804}. This model is widely applied to implement QST protocol in various setups such as opto-mechanical systems and macroscopic harmonic oscillators~\cite{RevModPhys.86.1391,LiOuLeiLiu+2021+2799+2832}. The Hamiltonian can be written in the interaction frame as (we set $\hbar=k_B = 1$ in this paper)
\begin{equation}
	\label{Hamiltonian_total}
	\tilde{H}_{\rm int}=\sum_{j=1}^{n}gk_j \left(a_jc^\dagger+ca_j^\dagger+a_jce^{-2i\omega t}+a_j^\dagger c^\dagger e^{2i\omega t} \right).
\end{equation}
The boson modes $a_j$ represents the $j$th oscillator-encoded qubits around the common central mode $c$. It's also assumed that all the modes are resonant with frequency $\omega$. The coupling strengths between the boson modes and the central mode can be effectively controlled by rectangle pulses focused on different mode, and the relative amplitude of the $j$th pulse is denoted by $k_j$. 

The multi-mode bosonic model can be realized in circuit quantum electrodynamical systems, such as superconductor chips~\cite{PhysRevX.7.011035,PRXQuantum.2.030321}, where spatially separated superconducting qubits are connected by inductor-capacitor (LC) circuit~\cite{RN82}. The LC circuit can serve as a single-mode quantum data bus with high quality factor for transferring quantum information between superconductor qubits. In addition, every superconductor qubit individually couples to a single-mode readout cavity. Thus, by connecting the $n$ readout cavities with a common LC circuit one can realize the model under consideration.

Distinguished from other approaches based on RWA in the long-time limit $t\gg1/\omega$, the counterrotation terms $a_j ce^{-2i\omega t}$ and $a_j^\dagger c^\dagger e^{2i\omega t}$  are retained in our model, thus breaking the global U$(1)$ symmetry. Such generalization makes it a more realistic model in the strong-coupling regime, or equivalently, short-time limit $t\sim1/\omega$. By solving the Heisenberg equation of the system, we can obtain an analytical expression of the linear transformation for all the boson modes as a function of evolution time $t$ (see \ref{App:solution}). 

In the following sections, we will first focus on a simple case of $n=2$ to demonstrate our designed fast linear bosonic transformation. In such three-mode (two boson modes plus one channel mode) system, the analytical solution can be obtained by defining a united mode $a(t)=\frac{k_1a_1(t)+k_2a_2(t)}{\sqrt{k_1^2+k_2^2}}$, which reads, 
\begin{eqnarray}
	a(t)&=&\left(
		e^{i\sqrt{\omega^2-2g\omega}t},\ e^{-i\sqrt{\omega^2-2g\omega}t},\ e^{i\sqrt{\omega^2+2g\omega}t}, \ e^{-i\sqrt{\omega^2+2g\omega}t}
	\right) M
\left(\begin{array}{c}
	a(0)\\a^\dagger(0)\\c(0)\\c^\dagger(0)
\end{array}\right) e^{i\omega t},\nonumber\\
M&=&\frac{1}{4}\left(\begin{array}{cccc}
	1-\frac{\zeta-1}{\sqrt{\zeta^2-2\zeta}}&\frac{1}{\sqrt{\zeta^2-2\zeta}}&-1+\frac{\zeta-1}{\sqrt{\zeta^2-2\zeta}}&-\frac{1}{\sqrt{\zeta^2-2\zeta}}\\
	1+\frac{\zeta-1}{\sqrt{\zeta^2-2\zeta}}&-\frac{1}{\sqrt{\zeta^2-2\zeta}}&-1-\frac{\zeta-1}{\sqrt{\zeta^2-2\zeta}}&\frac{1}{\sqrt{\zeta^2-2\zeta}}\\
	1-\frac{\zeta+1}{\sqrt{\zeta^2+2\zeta}}&-\frac{1}{\sqrt{\zeta^2+2\zeta}}&1-\frac{\zeta+1}{\sqrt{\zeta^2+2\zeta}}&-\frac{1}{\sqrt{\zeta^2+2\zeta}}\\
	1+\frac{\zeta+1}{\sqrt{\zeta^2+2\zeta}}&\frac{1}{\sqrt{\zeta^2+2\zeta}}&1+\frac{\zeta+1}{\sqrt{\zeta^2+2\zeta}}&\frac{1}{\sqrt{\zeta^2+2\zeta}}
\end{array}\right).
\end{eqnarray}
Here, $\zeta={\omega}/{g}$ stands for the inverse coupling strength. The solution for mode $c$ can be obtained analogously by exchanging $a$ and $c$ in the solution of mode $a$. Notice that in the case of $| g| >\omega/2$, we define $\sqrt{\omega^2\mp2g\omega}=i\sqrt{2g\omega\mp\omega^2}$ and $\sqrt{\zeta^2\mp2\zeta}=i\sqrt{2\zeta\mp\zeta^2}$ to avoid possible confusion of double-valued square root. 

Several conditions need to be considered to obtain the evolution of individual modes $a_1(t)$ and $a_2(t)$. The modes $a_1$ and $a_2$ still demonstrate symmetry in exchange of the coupling strengths $k_1$ and $k_2$. Besides, by defining a united mode $a$, an initial condition associated with the orthogonal combination of $a_1$ and $a_2$ is dropped out from the problem and needs to be reintroduced. With all these considerations, the solutions can be expressed as,
\begin{eqnarray}
	a_1(t)&=&\frac{k_1}{\sqrt{k_1^2+k_2^2}}a(t)+B, \nonumber \\
	a_2(t)&=&\frac{k_2}{\sqrt{k_1^2+k_2^2}}a(t)-\frac{k_1}{k_2}B,
	\label{eqn:solution}
\end{eqnarray}
where $B=a_1(0)-\frac{k_1}{k_1^2+k_2^2}[k_1a_1(0)+k_2a_2(0)]$ is determined by the initial condition. The solution of mode $c$ can be directly obtained according to the equivalent position between $c(t)$ and the united mode $a(t)$.

\section{QST of a single boson mode} 
\label{sec:QST}

In this section, we discuss the application of the analytic solution to tasks of transferring a single boson mode from the sender $a_1$ to the receiver $a_2$ through the intermediate channel mode $c$, i.e., taking the simplest example of $n=2$ in Eq.~(\ref{Hamiltonian_total}). For simplicity, we consider equal couplings $k_1=k_2=1$, while the qualitative conclusions can be generalized to other cases. In the weak-coupling regime, the implementation of QST has been studied using conventional method based on RWA~\cite{Parkins_1999,PhysRevA.50.R3589,PhysRevX.7.011035,PhysRevLett.118.133601,Xiao-Ling,PhysRevLett.106.040505,PhysRevA.87.012339}. By neglecting the counterrotation terms $a_j ce^{-2i\omega t}$ and $a_j^\dagger c^\dagger e^{2i\omega t}$ in the long-time limit $t\gg1/\omega$, the initial Hamiltonian is reduced to $\tilde{H}_{\rm int}=\sum_{j=1}^{2}g (a_j c^\dagger+ca_j^\dagger)$, with global U$(1)$ symmetry and conserved total excitation number. Then, the Heisenberg equations can be obtained as 
\begin{equation}
\dot{a}_{1(2)}(t)=-igc(t),\; \dot{c}(t)=-ig[a_1(t)+a_2(t)]. 
\end{equation}
Here the coupling strength $g$ and the duration $\tau$ of the rectangle pulse should satisfy the relation $g'\tau=\pi$, where $g'\equiv\sqrt{2}g$ is the effective coupling strength determined by the pulse amplitude~\cite{Parkins_1999,PhysRevA.50.R3589,PhysRevX.7.011035,Xiao-Ling,PhysRevLett.106.040505,PhysRevA.87.012339}. Under such condition, the exchange of modes $a_1$ and $a_2$ can be realized, $a_1(\tau)=-a_2(0)$, $a_2(\tau)=-a_1(0)$, while the intermediate mode $c$ remaining unchanged, $c(\tau)=-c(0)$. However, this perfect result of QST is only an artifact rooted from the assumed global U$(1)$ symmetry which is only approximately preserved for weak coupling.

\subsection{Optimized bosonic transformation for fast QST}

When aiming towards high-speed QST, a strong coupling is required and counterrotation terms become non-negligible, which break the global U$(1)$ gauge invariance and the conservation of total excitation number, and lead to sizable infidelity of the final state. Thus, one needs to go beyond RWA and work with the original Hamiltonian. 

Defining the vector of the operators as $\Psi=(a_1,c,a_2)^T$ and $\Psi^{d}=(a_1^{\dagger},c^{\dagger},a_2^{\dagger})^T$, the dynamics of all the three modes right after a duration $t$ can be expressed by the block matrix form in the Heisenberg picture,
\begin{eqnarray}
\label{vector expression}
    \left(\begin{array}{c}
          \Psi(t)\\ \Psi^d(t)  
    \end{array}\right)=\left(\begin{array}{cc}
        \mathcal{U}_A(t) & \mathcal{U}_B(t) \\
        \mathcal{U}_B^*(t) & \mathcal{U}_A^*(t)
    \end{array}\right)\left(\begin{array}{c}
          \Psi(0)\\ \Psi^d(0)  
    \end{array}\right).
\end{eqnarray}
Typically, the creation and annihilation vectors couple with each other by the off-diagonal transformation matrices $\mathcal{U}_B(t)$ and $\mathcal{U}_B^*(t)$, and cause the variation of total excitation number. We focus on the first line of the 6-dimensional linear expression of Eq.~(\ref{vector expression}), and write the final state of mode $a_1$ right after the pulse duration $\tau$.
\begin{eqnarray}
	a_1(\tau)&=&K_{11}a_1(0)+K_{21}a_2(0)+K_{c1}c(0) \nonumber \\
	&&+K_{12}a_1^\dagger(0)+K_{22}a_2^\dagger(0)+K_{c2}c^\dagger(0),
\end{eqnarray}
where the coefficients $K$'s can be expressed as functions of $\tau$ and $g'$ (see \ref{App:QST} for details). A successful QST with high fidelity corresponds to the case with $\left|K_{21}\right|\sim1$ and all other coefficients $ |K| \sim0$. At the same time, in despite of the breakdown of U$(1)$ gauge invariance, we find that the excitation number can be preserved in specific conditions with $K_{12}=K_{22}=K_{c2}=0$. The optimized scheme of QST can be obtained by solving these equations, leading to
\begin{eqnarray}
	\label{pulse ansatz condition}
		\zeta&\equiv& \omega/g' = 2\frac{(1-2/m)^2+1}{1-(1-2/m)^2},\nonumber\\
		\theta& \equiv& \omega \tau = m\pi\sqrt{\frac{(1-2/m)^2+1}{2}}.
\end{eqnarray}
Here, $m=2,3,\dots$ is an arbitrary positive integer. One can easily find that $\zeta$ and $\theta$ increase monotonously with parameter $m$. 
\begin{figure}[tb]
	\centering
	\includegraphics[width=1.0\textwidth]{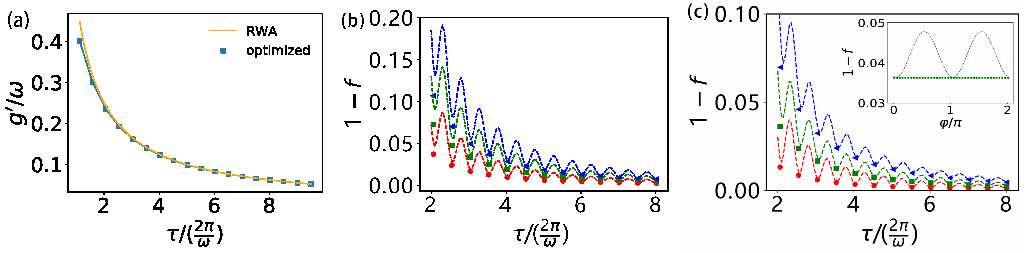}
	\caption{(a) The QST duration $\tau$ and coupling strength $g'$ determined by our optimized method and RWA. (b) The infidelity when transferring Fock states $\ket{n,0,0}\to\ket{0,0,n}$ with $n=1$ (red), $2$ (green), and $3$ (blue). (c) The infidelity when transferring coherent states $\ket{\alpha,0,0}\to\ket{0,0,\alpha}$ with $\alpha=0.6e^{i{\pi}/{2}}$ (red), $e^{i{\pi}/{2}}$ (green), $1.4e^{i{\pi}/{2}}$ (blue). Inset of (c) The infidelity changes with the phase $\varphi$ of the coherent state $\ket{\left|\alpha\right|e^{i\varphi}}$ with fixed amplitude $\alpha=1.0$, where the pulse parameter is set as $m=5$. In (b) and (c), the integer $m$ is taken from 5 to 17, corresponding to an evolution time $\tau$ from $2\times\frac{2\pi}{\omega}$ to $8\times\frac{2\pi}{\omega}$. The solid dots and dashed lines stand for the optimized method and RWA method, respectively. The analytical prediction of the error Eq.~(\ref{function relation}) agrees well with the solid dots, being indistinguishable in this plot (see \ref{App:QSL}).}
	\label{comparison with RWA}
\end{figure}

The condition Eq.~(\ref{pulse ansatz condition}) provides an implicit constraint to optimize the QST procedure. First, one needs to choose a suitable parameter $m$. The amplitude and duration of the rectangle pulse are then determined by $\zeta$ and $\theta$, respectively, both acquiring discrete values. As shown in Fig.~\ref{comparison with RWA}(a), the optimized pulse parameters agree well with those obtained within RWA in the weak-coupling limit of small $g^\prime$, but deviate in the strong-coupling regime. Under the optimized pulse condition, the final mode can be simplified as $a_1(\tau)=K_{11}a_1(0)+K_{21}a_2(0)$ (see \ref{App:QST}), where
\begin{eqnarray}
    \label{K_11}
	K_{11}&=&\sin(\theta_r)e^{i(\theta_r-{\pi}/{2})},\nonumber \\
	\label{K_21}
	K_{21}&=&-\cos(\theta_r)e^{i\theta_r},\nonumber \\
	\label{rotation angle}
	\theta_r&\equiv&-\frac{1}{4} (\sqrt{1+{2}/{\zeta}}+\sqrt{1-{2}/{\zeta}}-2)\theta.
\end{eqnarray}
This result suggests that both the amplitude error and phase error of QST are dependent on a single parameter, the phase shift $\theta_r$, which is analytically determined by the pulse parameters and can be compensated by a local rotation as we discuss latter. With this pulse ansatz, the annihilation and creation operators are decoupled temporarily, i.e. $\mathcal{U}_B(\tau)=0$, indicating that the total excitation number is conserved at this point although the global U$(1)$ symmetry is broken. A detailed discussion is provided in \ref{App:conservation}. The linear transformation of the bosonic operators takes the form
\begin{eqnarray}
\Psi(\tau)=\mathcal{U}_A(\tau)\Psi(0),
\end{eqnarray}
with transform matrix 
\begin{eqnarray}
\mathcal{U}_A(\tau)=\left(\begin{array}{ccc}
  K_{11} & 0 & K_{21} \\
  0 & -e^{2i\theta_r} & 0 \\
  K_{21} & 0 & K_{11} \\
\end{array}\right)\approx\left(\begin{array}{ccc}
  0 & 0 & -1 \\
  0 & -e^{2i\theta_r} & 0 \\
  -1 & 0 & 0 \\
\end{array}\right),
\end{eqnarray}
making it possible for high-fidelity QST. 

To demonstrate the performance of the optimized scheme, we consider a QST task from the initial state $\ket{\Psi(0)}=\ket{\psi,0,0}$ to the goal state $\ket{\Psi_{\rm goal}}=\ket{0,0,\psi}$. The three parts in Dirac kets represent the state in node $a_1$ (sender), channel $c$ and node $a_2$ (receiver), respectively. As an example, we first demonstrate the results of transferring a Fock state $\ket{\psi}=\ket{n}$ with $n=1,2,3$ [Fig.~\ref{comparison with RWA}(b)], and transferring a coherent state $\ket{\psi}=\ket{\alpha}$ with $\alpha=0.6e^{i{\pi}/{2}},e^{i{\pi}/{2}},1.4e^{i{\pi}/{2}}$ [Fig.~\ref{comparison with RWA}(c)] through an ideal channel at zero temperature. By choosing the parameter $m$ from $5$ to $17$, the infidelity of the received state $1-f$ changing with the pulse duration $\tau$ is presented, where the fidelity $f\equiv {\rm tr}(\sqrt{\sqrt{\rho_i}\rho_f\sqrt{\rho_i}})^2= {\rm tr}(\rho_{f}\rho_i)$ with $\rho_{i}=\ket{\psi}\bra{\psi}$ being the density matrix of the sending state at node $a_1$ and $\rho_{f}\equiv {\rm tr}_{1,c}[\rho_{1,c,2}(\tau)]$ the reduced density matrix of the received state at node $a_2$~\cite{jozsa1994fidelity}. The optimized pulse given by Eq.~(\ref{pulse ansatz condition}) outperforms the one predicted by RWA for all cases. In the weak-coupling regime $g'\lesssim0.1\omega$, where the parameter is chosen as $m \gtrsim 11$ and the pulse duration $\tau \gtrsim 5\times ({2\pi}/{\omega})$, the optimized scheme reduces to the pulse ansatz under RWA. In the strong-coupling regime $g' \gtrsim0.1 \omega$ with $m \lesssim 11$, our scheme significantly suppresses infidelity from the RWA result, which presents significant fluctuation owing to the non-negligible effect of the counterrotation terms. The reduction of infidelity over the best performance of RWA can be as large as $\sim$20\%, which is substantial considering the already low enough baseline. The ultra strong-coupling regime with $m=2,3,4$ is not shown in Fig.~\ref{comparison with RWA} since the infidelity will be too large to qualify a successful QST process. 

Another advantage of our optimized scheme is the independence on the phase of initial state when transferring coherent states. For an initial state $\rho=\ket{\alpha}\bra{\alpha}$ with $\alpha=\left|\alpha\right|e^{i\varphi}$, the infidelity will not change with the phase $\varphi$ as shown in the inset of Fig.~\ref{comparison with RWA}(c). As a comparison, for the RWA scheme, the infidelity oscillates with $\varphi$. This is because in RWA method, the final mode $a_1^\dagger(\tau)$ contains not only the target mode $a_2^\dagger(0)$ but also a portion of mode $a_2(0)$. The additional mode will result in a state similar to the target coherent state $\ket{\alpha}$ but with a loss of two photons in every Fock basis, which is expressed as $\ket{\psi}_{\rm err}=i\mathcal{E}e^{-{\left|\alpha\right|^2}/{2}}\sum_{n=2}^{\infty}\varepsilon_n\frac{\alpha^n}{\sqrt{n!}}\ket{n-2}$. The coefficient $i\mathcal{E}$ is pure imaginary with fixed phase $\pi/2$, and the real coefficient $\varepsilon_n$ in the summation is obtained from the binomial expansion $[K_{21}a_2^\dagger(0)+K_{22}a_2(0)+\dots]^n$. By combining the effects of these two modes, a relative phase of $\delta\theta=2\varphi+\pi/2$ will emerge and contribute a periodic modulation of infidelity. As a comparison, in our optimized method with the conditions (\ref{pulse ansatz condition}) satisfied, the final mode $a_1^\dagger(\tau)$ only contains the target mode $a_2^\dagger(0)$, such that the infidelity is not affected by the phase $\varphi$.

\subsection{Tradeoff between speed and fidelity} 
One key merit of having the analytic solution Eq.~(\ref{K_11}) is that the explicit relation between transfer speed and fidelity of QST can be obtained for the optimized scheme of rectangle pulses. We derive a general constraint of the initial pure state $\rho$ to be transferred, the evolution time $\tau$, and the infidelity $1-f$, in the form of
\begin{eqnarray}
	\label{function relation}
	F(\rho,1-f,\tau)=1-f-\left<n_1\right>_\rho\left|K_{11}(t=\tau)\right|^2 = 0.
\end{eqnarray}
According to Eq.~(\ref{K_11}), we can define the magnitude of coefficient $G(m) \equiv |K_{11}|$ as 
\begin{eqnarray}
	G(m) = \sin\left[-\left(\sqrt{1+\frac{2}{\zeta(m)}}+\sqrt{1-\frac{2}{\zeta(m)}}-2 \right) \frac{\theta(m)}{4}\right],
\end{eqnarray}
where $m$ is taken as a continuous variable temporarily. For a task of sending state $\rho$ with an upper bound of tolerable error $E_{\rm tol} \ge 1-f$, the QST time $\tau$ must satisfy $G(m) \le \sqrt{{E_{\rm tol}}/{\left<n_1\right>_\rho}}$. By solving the inverse function of $G(m)$, we can get the threshold (lower bound of $m$) $m_{\rm th}\equiv G^{-1}(\sqrt{{E_{\rm tol}}/{\left<n_1\right>_\rho}})$. Thus, the threshold of $\theta$ can be obtained by $\theta_{\rm th}=\theta(\lfloor m_{\rm th}\rfloor+1)$, where $\lfloor m_{\rm th}\rfloor$ is the nearest integer less than or equal to $m_{\rm th}$. The fastest possible time to accomplish the transfer, i.e., the quantum speed limit of QST, is $\tau_{\rm th}={\theta_{\rm th}}/{\omega}$. We demonstrate a detailed comparison between the infidelity in our optimized method and that predicted by the tradeoff relation in Fig.~\ref{QSL} in \ref{App:QSL}, which shows excellent agreement for all parameters.

\subsection{Robustness against fluctuations} 

To demonstrate the feasibility of the proposed scheme under realistic experimental conditions, in the following we investigate the thermal noise effect of the channel, and the imperfection of pulse shape. 
\begin{figure}[t]
	\centering
	\includegraphics[width=0.8\textwidth]{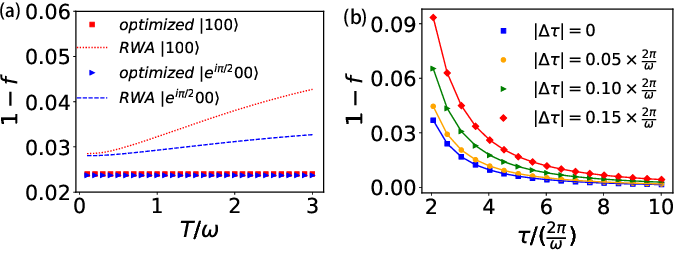}
	\caption{(a) The infidelity versus channel temperature $T$ for tasks of transferring the Fock state $\ket{1}$ (red) and the coherent state $\ket{e^{i{\pi}/{2}}}$ (blue) by the optimized and RWA methods, respectively. The pulse parameter is chosen as $m=6$. (b) The infidelity for transferring the Fock state $\ket{1}$ with different pulse duration fluctuations $|\Delta \tau|$ in the optimized pulse scheme. Here, an ideal channel with $T=0$ is assumed, and solid lines as guides for the eyes are drawn to connect symbols.}
	\label{robust}
\end{figure}

For the thermal noise, suppose the initial state of the channel is a thermal state $\rho_c=e^{-\beta H_c}/Z$ rather than a vacuum state at zero temperature, where $\beta={1}/{T}$ is the inverse temperature, $Z={\rm tr}(e^{-\beta H_c})$ is the partition function of Maxwell-Boltzmann distribution, and $H_c$ is the channel mode Hamiltonian. In Fig.~\ref{robust}(a), we compare the optimized scheme with RWA method for temperature up to $T\sim 3/\omega$. While the infidelity of the RWA scheme increases almost linearly with temperature, and showing a worse performance when transferring a Fock state than a coherent state, the optimized method is completely immune to thermal noise in both cases. To understand this observation, we consider as an example the task of transferring an initial Fock state $| n_1 \rangle$ through a channel with an initial phonon number $n_c$. Note that at the end of the pulse duration $\tau$, the final state of the nodes are completely decoupled from the channel, leading to $a_1(\tau)=K_{11}a_1(0)+K_{21}a_2(0)$, $a_2(\tau)=K_{11}a_2(0)+K_{21}a_1(0)$, and $c(\tau)=-e^{2i\theta_r}c(0)$. The final state is given by
\begin{eqnarray}
	\ket{\Psi(\tau)}&=&U(\tau)\frac{(a_1^\dagger)^{n_1}}{\sqrt{n_1!}}\frac{(c^\dagger)^{n_c}}{\sqrt{n_c!}}\ket{0_1,0_c,0_2}\nonumber\\
	&=&\frac{(U(\tau)a_1^\dagger U^\dagger(\tau))^{n_1}}{\sqrt{n_1!}}\frac{(U(\tau)c^\dagger U^\dagger(\tau))^{n_c}}{\sqrt{n_c!}}U(\tau)\ket{0_1,0_c,0_2}\nonumber\\
	&=& \frac{(K_{11}a_1^\dagger+K_{21}a_2^\dagger)^{n_1}}{\sqrt{n_1!}}\frac{(-e^{-2i\theta_r}c^\dagger)^{n_c}}{\sqrt{n_c!}}\ket{0_1,0_c,0_2}\nonumber\\
	&=&\frac{(K_{11}a_1^\dagger+K_{21}a_2^\dagger)^{n_1}}{\sqrt{n_1!}}\ket{0_1,n_c,0_2}.
\end{eqnarray}
Here, $U(\tau)$ is the unitary operator describing the evolution of the system with time $\tau$. This result shows that after the pulse is applied, the channel mode evolves back to its initial state, being independent on the status of nodes $a_{1,2}(0)$. Meanwhile, the influence of the channel on the nodes is also erased at this exact time after a partial trace. This conclusion can be easily generalized to a thermal channel, which is a classical superposition of Fock states with different $n_c$, and also to the case of transferring a coherent state as it can be expanded into Fock basis.

For fluctuation effect induced by the pulse, we stress that in the more interesting strong-coupling regime with large pulse amplitude and short duration, a small deviation of coupling intensity $\Delta g$ should be less significant in comparison to an error of duration $\Delta \tau$. In Fig.~\ref{robust}(b), we impose different amount of $\Delta \tau$, and plot the most prominent infidelity $1-f$ for a pulse duration within $[\tau - \Delta \tau, \tau+ \Delta \tau]$. Here, the fidelities for different $|\Delta \tau|$ are displayed by symbols, which are connected by solid lines as a guide for the eyes. As expected, the fluctuation of pulse time is more influential for stronger coupling and shorter transfer time. However, if the fluctuation is relatively small with $\Delta \tau\le0.05\times2\pi/\omega$, our optimized method still performs well and the increase of infidelity is restricted within one percent even for the strongest coupling considered.

\subsection{Correction of the phase error} 
Based on the analytic bosonic transformation, we can obtain the phase rotation of the final mode as in Eq.~(\ref{rotation angle}). This effect is more severe in the strong-coupling regime, which induces a sizable error to the final state. With that knowledge, one can further enhance the fidelity of QST by applying a local rotation to node $a_2$. It can be realized by implementing a local pulse on node $a_2$ after the coupling pulse (\ref{pulse ansatz condition}), which is expressed as $H_r=g_r(t)a_2^\dagger a_2$ with the pulse area $\int g_r(t)dt=\theta_r$. Then the corrected bosonic vector can be expressed as $\Psi_r=\mathcal{U}_r\Psi(\tau)$, where $\mathcal{U}_r={\rm diag}[1,1,e^{-i\theta_r}]$. As shown in Fig.~\ref{suppfig(f)} in \ref{App:rotation}, such a simple local rotation improves the transferring fidelity remarkably for a coherent state. However, this method does not make any difference for a Fock state since the Wigner function of which is central symmetric. 
\begin{figure}[tb]
	\centering
	\includegraphics[width=\textwidth]{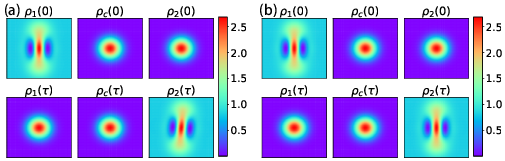}
	\caption{The initial (top row) and final (bottom row) states of a QST of cat state $\ket{{\rm cat} (\alpha)}_{+}$ with $\alpha=1.2$ through the optimized method. The pulse parameter is chosen as $m=11$. (a) When the local rotation is not applied, the fidelity of the final state is $f=0.9819$. (b) After the local rotation, the fidelity is increased to $f=0.9922$.}
	\label{rotation remedy}
\end{figure}

To elucidate it more explicitly, we consider a task of transferring an even cat state in continuous variable systems $\ket{{\rm cat}(\alpha)}_+\equiv\ket{\alpha}+\ket{-\alpha}$ through an ideal zero temperature channel. As a typical macroscopic quantum superposition, cat state has been prepared in various physical platforms including optical systems~\cite{ourjoumtsev2006generating,PhysRevLett.101.233605,han2023remote}, opto-mechanical systems~\cite{PhysRevLett.116.163602}, superconducting systems~\cite{Yang:18}, atomic ensembles~\cite{doi:10.1126/science.272.5265.1131,doi:10.1126/science.aay0600} and magnon-photon systems~\cite{PhysRevLett.127.087203}. In Fig.~\ref{rotation remedy}(a), we apply our optimized method directly and then observe the final state in node $a_2$, which presenting a rotation of an extra phase as expected. In Fig.~\ref{rotation remedy}(b), we apply the local rotation to node $a_2$ right after the coupling pulse (\ref{pulse ansatz condition}), and the fidelity is obviously increased.

\section{QST for multi-mode W-type state} 

Next, we show that the protocol developed for single-mode QST can be extended for transferring multi-mode quantum states. For definiteness, we consider as an example an $n_s$-mode W-type state, which is denoted by $\ket{\psi_s}=\sum_{j=1}^{n_s}C_j\ket{1_{j}}\otimes\ket{0_{\bar{j}}}$ with $\ket{0_{\bar{j}}}\equiv\otimes_{i\ne j}\ket{0_i}$. Without loss of generality, we assume all coefficients $C_j$ ($j=1,2,\dots,n_s$) are real, since a complex amplitude can be realized by applying corresponding local phase gate or equivalently by a proper redefinition of the corresponding boson mode  $a_j \rightarrow a_j e^{i\phi}$. The initial state of the whole system can be expressed as $\ket{\psi(0)}_{sr}=\ket{\psi_s}\otimes\ket{0_r}$, where the receiver side is assumed to be an $n_s$-mode vacuum state. In the following, we denote the $n_s$ boson modes in the sender side as $a_1,\dots,a_{n_s}$, and the $n_s$ boson modes in the receiver side as $a_{n_s+1},\dots,a_{2n_s}$ to simplify notation. Notice that since the initial state of the channel mode $\ket{\psi_c}$ will have no influence on the final state, we can trace out this degree of freedom and focus on the $2n_s$ boson modes in sender and receives sides only. Using this notation, the initial and final states can be obtained as 
\begin{equation}
	\label{final multi-mode state}
	\ket{\psi(0)}_{sr}=\sum_{j=1}^{n_s}C_ja_j^\dagger(0)\ket{0}_{sr}\rightarrow\ket{\psi(\tau)}_{sr}=\sum_{j=1}^{n_s}C_ja_j^\dagger(\tau)\ket{0}_{sr},
\end{equation}
where $\ket{0}_{sr} = \ket{0}_{s} \otimes \ket{0}_{r}$ denotes the vacuum state of both the sender and receiver.

By using the complete expression in \ref{App:multimode}, an exact result of the final state can be written down and an optimized transfer scheme can be obtained. On the other hand, to demonstrate the feasibility and advantage of our proposed scheme, in the following discussion we ignore the error caused by the strong coupling temporarily, and write the approximate linear bosonic transformation for the $2n_s$ modes system as
\begin{equation}
	a_i(\tau)\approx a_i(0)-\frac{2k_i}{\sum_{j=1}^{2n_s}k_j^2}\sum_{j=1}^{2n_s}k_ja_j(0).
\end{equation}
Combining with the expression of Eq.~(\ref{final multi-mode state}), the final state can be expanded with the Fock basis of initial boson modes, leading to
\begin{equation}
	\ket{\psi(\tau)}_{sr}=\sum_{i=1}^{n_s}C_i \left[a_i^\dagger(0)-\frac{2k_i}{\sum_{j=1}^{2n_s}k_j^2}\sum_{j=1}^{2n_s}k_ja_j^\dagger(0) \right]\ket{0}_{sr}.
\end{equation}
The analytic expressions of the coefficients of the $2n_s$ modes thus allow us to realized QST by properly designing the coupling strengths $k_j$'s.

Firstly, we demand that excitations should not occur in the sender side. This requirement gives a total of $n_s$ restrictions for the $n_s$ coefficients of the sender modes 
\begin{equation}
	\sum_{i=1}^{n_s}C_i \left( \delta_{ij_0}-\frac{2k_ik_{j_0}}{\sum_{j=1}^{2n_s}k_j^2} \right)=0,
\end{equation}
where $j_0=1,2,\dots,n_s$. Then, the relative coupling strengths in the receiver side should be designed in a certain proportion according to the corresponding probability amplitudes of the initial W-type state. This gives another $n_s-1$ conditions
\begin{equation}
	\frac{\sum_{i=1}^{n_s}C_ik_ik_{n_s+j_1}}{\sum_{i=1}^{n_s}C_ik_ik_{n_s+j_2}}=\frac{k_{n_s+j_1}}{k_{n_s+j_2}}=\frac{C_{j_1}}{C_{j_2}},
\end{equation}
where the index $j_1$ and $j_2$ can be arbitrary chosen from $1$ to $n_s$.

Taking $n_s=2$ as an example, the initial two-mode entangled state (W-type state) takes the form as $\ket{\psi_s}=C_1\ket{10}+C_2\ket{01}$. So the two restrictions on the relative coupling strengths in the sender side read
\begin{eqnarray}
	\label{W-state coupling restrict}
	C_1\left(1-\frac{2k_1^2}{\sum_{j=1}^{4}k_j^2}\right)+C_2\left(-\frac{2k_1k_2}{\sum_{j=1}^{4}k_j^2}\right)&=0\\
	C_1\left(-\frac{2k_1k_2}{\sum_{j=1}^{4}k_j^2}\right)+C_2\left(1-\frac{2k_2^2}{\sum_{j=1}^{4}k_j^2}\right)&=0.
\end{eqnarray}
The relation between $k_1$ and $k_2$ thus can be derived
\begin{equation}
	k_1^2=k_2^2+\frac{(C_1^2-C_2^2)(k_3^2+k_4^2)}{C_1^2+C_2^2}.
\end{equation}
Next, by substituting the expression above of $k_1$ into Eq.~(\ref{W-state coupling restrict}), we can get the equation for $k_2$ 
\begin{equation}
	\left(k_2^2\right)^2+\frac{C_1^2-C_2^2}{C_1^2+C_2^2}\left(k_3^2+k_4^2\right)k_2^2- \left[\frac{C_1C_2}{C_1^2+C_2^2}(k_3^2+k_4^2) \right]^2=0.
\end{equation}
There must exist a positive solution for $k_2^2$. So we can determine the value for the coupling strengths in the sender side to achieve a successful transfer of multi-mode W-type states with high fidelity. For more general multi-mode states, the possible existence of multi-excitations makes QST a doable task in principle, but a harder challenge in practice. One possible solution is to go beyond the framework of time-independent rectangle pulse control, and combine our complete dynamic expression in \ref{App:multimode} with the time-dependent pulse optimization method~\cite{PhysRevLett.103.240501,PhysRevLett.118.150503}.

\section{Entanglement Preparation}
\label{sec:EP}

Quantum entanglement serves as a fundamental element for quantum computation and quantum communication~\cite{RN90,RN91,doi:10.1126/science.aaw9415}. The generation and distribution of entangled states have long been the main topic in the field of quantum information processing~\cite{PRXQuantum.2.030321,RN97,RN98,RN100,RN101,PhysRevApplied.17.054021}. In linear optical systems, entanglement can be prepared by certain bosonic transformation such as beam splitter. Making use of the result of linear bosonic transformation obtained before, we raise a new approach to generate multi-mode entangled states in strongly coupled bosonic systems, and give the shortest possible generation time under our protocol.

We first consider the simplest three-mode model with two boson modes $a_{1,2}$ coupled to a single-mode channel $c$. According to the dynamic solutions Eq.~(\ref{eqn:solution}), the intermediate mode $c$ at an arbitrary evolution time $t$ reads
\begin{eqnarray}
	\label{c solution}
	c(t)&=&\Bigg\{ \frac{c(0)}{2} \bigg[\cos \left(\sqrt{\omega^2-2g'\omega}t \right)+\cos \left(\sqrt{\omega^2+2g'\omega}t \right)\nonumber\\
	&& \hspace{1cm} -i\frac{\zeta-1}{\sqrt{\zeta^2-2\zeta}}\sin \left(\sqrt{\omega^2-2g'\omega}t \right)-i\frac{\zeta+1}{\sqrt{\zeta^2+2\zeta}}\sin \left(\sqrt{\omega^2+2g'\omega}t \right) \bigg]\nonumber\\
	&&+\frac{k_1a_1(0)+k_2a_2(0)}{2\sqrt{k_1^2+k_2^2}} \bigg[\cos \left(\sqrt{\omega^2+2g'\omega}t \right)-\cos \left(\sqrt{\omega^2-2g'\omega}t \right)\nonumber\\
	&& \hspace{1cm} +i\frac{\zeta-1}{\sqrt{\zeta^2-2\zeta}}\sin \left(\sqrt{\omega^2-2g'\omega}t \right)-i\frac{\zeta+1}{\sqrt{\zeta^2+2\zeta}}\sin \left(\sqrt{\omega^2+2g'\omega}t \right) \bigg]\nonumber\\
	&&+\frac{c^\dagger(0)}{2} \left[\frac{i}{\sqrt{\zeta^2-2\zeta}}\sin(\sqrt{\omega^2-2g'\omega}t)-\frac{i}{\sqrt{\zeta^2+2\zeta}}\sin \left(\sqrt{\omega^2+2g'\omega}t \right) \right]\nonumber\\
	&&+\frac{k_1a_1^\dagger(0)+k_2a_2^\dagger(0)}{2\sqrt{k_1^2+k_2^2}} \left[-\frac{i}{\sqrt{\zeta^2-2\zeta}}\sin \left(\sqrt{\omega^2-2g'\omega}t \right)\right.\nonumber\\
	&&\hspace{1cm}\left.-\frac{i}{\sqrt{\zeta^2+2\zeta}}\sin \left(\sqrt{\omega^2+2g'\omega}t \right) \right] \Bigg\}e^{i\omega t},
\end{eqnarray}
where $k_{1,2}$ denote the coupling weights between modes $a_{1,2}$ and mode $c$. Similarly, in order to avoid the error introduced by strong coupling, the condition of conserved total excitations at discrete times is assumed to be satisfied
\begin{eqnarray}
	\label{optimized pulse ansatz_entanglement}
		&&\left(\sqrt{1+\frac{2}{\zeta}}-\sqrt{1-\frac{2}{\zeta}}\right) \theta=\pi,\nonumber\\
		&&\sqrt{1+\frac{2}{\zeta}}\theta=m\pi.
\end{eqnarray}
Here, $m=2,3,\dots$ is an arbitrary positive integer. We can solve the above constraints analytically and give the similar pulse condition for EP tasks,
\begin{eqnarray}
	\label{optimized pulse ansatz_entanglement 2}
		&&\zeta=2\frac{(1-1/m)^2+1}{1-(1-1/m)^2},\nonumber\\
		&&\theta=m\pi\sqrt{\frac{(1-1/m)^2+1}{2}}.
\end{eqnarray}
The time $\tau$ and the effective coupling strength $g'$ can be easily obtained from Eq.~(\ref{optimized pulse ansatz_entanglement 2}) as $\theta\equiv\omega \tau$ and $\zeta\equiv\omega/g'$. By substituting these solutions into Eq.~(\ref{c solution}), we can simplify the expression of mode $c$,
\begin{equation}
	c(\tau)=(-1)^m\frac{k_1a_1(0)+k_2a_2(0)}{\sqrt{k_1^2+k_2^2}}e^{i\theta}.
\end{equation}

Notice that the pulse ansatz Eq.~(\ref{optimized pulse ansatz_entanglement 2}) is similar to that of QST Eq.~(\ref{pulse ansatz condition}) by scaling the parameter $m \to m/2$. In fact, if one sets an optimized pulse amplitude for EP, the amplitude of mode $c(0)$ is distributed into modes $a_1(\tau)$ and $a_2(\tau)$, and the state of mode $a_{1,2}(0)$ is transferred to the intermediate mode $c(\tau)$ after an optimized duration of EP. By doing so, an EP task is accomplished. If one fixes the coupling strength and extends the time by another EP pulse duration, the amplitudes in modes $a_1(\tau)$ and $a_2(\tau)$ will return back to mode $c(2\tau)$ gradually, and the initial state of modes $a_{1,2}(0)$ stored in the intermediate mode $c(\tau)$ will transfer to $a_{2,1}(2\tau)$. Then a QST task is completed. In addition, similar to the QST cases, the creation and annihilation operators are also decoupled under this condition, since the compact linear transformation $\Psi(\tau)=\mathcal{U}_A(\tau)\Psi(0)$ is also valid with 
\begin{eqnarray}
    \mathcal{U}_A(\tau)=\left(\begin{array}{ccc}
   \frac{k_2^2}{k_1^2+k_2^2} & \frac{k_1}{\sqrt{k_1^2+k_2^2}}& -\frac{k_1k_2}{k_1^2+k_2^2} \\
   -\frac{k_1}{\sqrt{k_1^2+k_2^2}} &0 &-\frac{k_2}{\sqrt{k_1^2+k_2^2}} \\
   -\frac{k_1k_2}{k_1^2+k_2^2} &\frac{k_2}{\sqrt{k_1^2+k_2^2}} & \frac{k_1^2}{k_1^2+k_2^2} \\
\end{array}\right).
\end{eqnarray}

Assuming that we encode a Fock state $\ket{1}$ on mode $c$, and leave modes $a_1$ and $a_2$ in the vacuum states $\ket{0}$, it is easily checked that the final state of modes $a_1$ and $a_2$ after one pulse of Eq.~(\ref{optimized pulse ansatz_entanglement 2}) is $\ket{\psi_{f}}=\frac{k_1\ket{1_10_2}+k_2\ket{0_11_2}}{\sqrt{k_1^2+k_2^2}}$. Thus, the states with different degrees of entanglement can be obtained by varying the relative coupling weights $k_1/k_2$. Specially, if the coupling weights are set as $k_1=\pm k_2$, a Bell state $\ket{\psi_{\rm Bell}}=(\ket{1_10_2}\pm\ket{0_11_2})/\sqrt{2}$ can be generated in a fast speed between modes $a_1$ and $a_2$ in the strong-coupling regime. 

\begin{figure}[tb]
	\centering
	\includegraphics[width=0.8\textwidth]{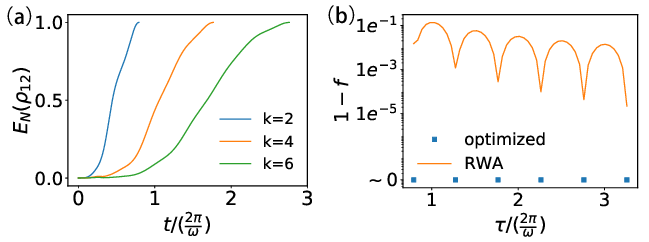}
	\caption{Generation of the maximal entangled state in nodes 1 and 2 with equal coupling strengths $k_1=k_2$. (a) Time evolution of log negativity between modes $a_{1}$ and $a_{2}$ under a rectangle-shaped pulse for $m=2,4,6$. (b) Infidelity between the final state and a perfect Bell state. Comparison is made between the optimized method (blue dots) and the RWA method (red line) for $m=2,3,\dots,7$.}
	\label{entanglement distribution}
\end{figure}

Then, we introduce the logarithmic negativity $E_N(\rho)\equiv {\rm log}_2(\left|\left|\rho^T\right|\right|_1)$ to quantify the degree of entanglement of the final state~\cite{PhysRevA.65.032314}, where the superscript $T$ means partial transportation of the $2$-mode density matrix and $\left|\left|\cdot\right|\right|_1$ stands for the trace norm of matrix. The maximum of logarithmic negativity for a 2-qubit system is $E_N(\rho)=1$, which is achieved for Bell states. In Fig.~\ref{entanglement distribution}(a) we display the evolution of logarithmic negativity of modes $a_1$ and $a_2$ for $m=2$, $4$, and $6$, with equal coupling strengths $k_1=k_2$. It is shown that a maximally entangled state will be prepared after applying the pulse introduced in Eq.~(\ref{optimized pulse ansatz_entanglement 2}). Besides, we also define the fidelity between the final state $\rho_f$ and the Bell state $\ket{\psi_{\rm Bell}}$, $f\equiv {\rm tr}[\ket{\psi_{\rm Bell}}\bra{\psi_{\rm Bell}} {\rm tr}_c(\rho_f)]$, and show the results of $1-f$ for $m= 2,3,\dots, 7$ in Fig.~\ref{entanglement distribution}(b). It is obvious that with our optimized pulse ansatz, the EP fidelity is always equal to $1$ for arbitrary cases with $m>1$. For comparison, the RWA method always gives a non-negligible infidelity, which can reach as high as $1\%$ in the strong-coupling regime. Thus, our optimized method provide a promising EP scheme to prepare perfect entangled states with low time cost. 

Remarkably, considering that in the optimized scheme the EP time is monotonically decreased with increasing coupling strength (and thus a smaller $m$), we conclude that the fastest time possible to generate a perfect entanglement can be obtained by setting $m=2$, which gives $\zeta=\omega/g'=\frac{10}{3}$, an $\theta=\omega\tau=\frac{\sqrt{10}}{2}\pi$. An even smaller choice of $m<2$ corresponds to an unbounded potential trap and the sine functions with imaginary variant in Eq.~(\ref{c solution}) are replaced by hyperbolic sine functions, leading to a diverging excitation number and an inevitable error (see \ref{App:QST} for details). 

Next, we show the above method can be directly generalized to the preparation of $N$-mode entangled states from $a_{1,2,\dots, N}$, where the evolution of the intermediate mode $c$ can be expressed as
\begin{equation}
	c(\tau)=(-1)^m\frac{\sum_{j=1}^{N}k_ja_j(0)}{\sqrt{\sum_{j=1}^{N}k_j^2}}e^{i\theta}.
\end{equation}
By defining the united mode $a(t) \equiv \frac{\sum_{j=1}^n k_j a_j(t)}{\sqrt{\sum_{j=1}^n k_j^2}}$ as in \ref{App:multimode}, the multi-mode EP scheme can be viewed as a linear transformation between $c(t)$ and the boson modes $a(t)$'s, expressed as 
\begin{eqnarray}
    \left(\begin{array}{c}
         a(\tau)\\c(\tau)
    \end{array}\right)=\left(\begin{array}{cc}
        0 & (-1)^me^{i\theta} \\
        (-1)^me^{i\theta} & 0
    \end{array}\right)
    \left(\begin{array}{c}
         a(0)\\c(0)
    \end{array}\right).
    \label{bosonic transformation for EP}
\end{eqnarray}
Thus, the $N$-mode entangled state can be obtained as
\begin{equation}
	\ket{\psi_{f}(k_1,k_2,\dots,k_N)}=\frac{\sum_{j=1}^{N}k_j\ket{1_j}\otimes\ket{0_{\bar{j}}}}{\sqrt{\sum_{j=1}^{N}k_j^2}},
\end{equation}
where $\ket{0_{\bar{j}}}\equiv\otimes_{i\ne j}\ket{0_i}$. Specifically, if the amplitudes $k_j$ are set equal, an $N$-mode W-type state \cite{PhysRevA.50.R2799} $\ket{\psi_{W}}=\sum_{j=1}^{N}\ket{1_j}\otimes\ket{0_{\bar{j}}}/\sqrt{N}$ is achieved. In addition, we notice that the proposed $N$-mode linear bosonic transformation for EP shown in Eq.~(\ref{bosonic transformation for EP}) takes the same form as a two-mode EP process except a different definition of $a(t)$. Thus, we conclude that the minimum EP time is independent on the number $N$ of modes.

\section{Summary} 
\label{sec:conclusion}

We introduce a universal optimized strategy for realizing fast linear bosonic transformation in the strong-coupling regime, where the creation and annihilation subspaces are decoupled, thus preserving the total excitation number and suppressing the infidelity of various quantum information tasks. Based on this strategy, we obtain the optimized pulse ansatz for quantum state transfer (QST) and entanglement preparation (EP), and demonstrate that both tasks can be achieved with high fidelity and fast speed. 

Firstly, we simulate the QST task between two boson modes coupled through a single-mode intermediate channel with rectangle shaped pulses, and demonstrate a reduction of infidelity up to $20$\% in the strong-coupling regime. By analytically solving the system, we obtain a tradeoff relation between the transferring speed and the tolerable error. This result can facilitate the choice of pulse amplitude and duration in experiment to optimally balance the QST time and fidelity. The proposed scheme is completely immune to thermal noise of the channel, and is robust against fluctuations of pulses. It can be further improved to approach higher fidelity by applying a local rotation of the final state to compensate the phase shift induced by the strong coupling, which is also universal and can be obtained analytically. Then we generalize the proposed method to QST of an arbitrary multi-mode W-type state through the common channel mode, demonstrating the potential to extend our protocol to more general quantum information tasks involving multi-mode dynamic processes. Secondly, we apply the optimized method to EP and propose a new approach to prepare multi-mode entangled states including Bell states and W-type states. The fastest preparation time based on our scheme has been obtained, which is independent on the number of the entangled modes. Our results provide new possibilities to achieve high fidelity and fast speed in QST and EP with realistic experimental techniques, and help reduce the stringent requirement of coherence time and temperature that would otherwise be required in quantum computation and quantum communication. In addition, the method may be combined with pulse optimization approach~\cite{PhysRevLett.103.240501,PhysRevLett.118.150503} to give even better pulse design. 

\ack
This work is supported by the National Natural Science Foundation of China (Grants No.~11975026, 12125402, 12147148, 12074428, and 92265208), and the National Key R\&D Program of China (Grants No.~2018YFA0306501 and No.~2022YFA1405301). F.-X. S. acknowledges the China Postdoctoral Science Foundation (Grant No.~2020M680186). Q. H. acknowledges the Innovation Program for Quantum Science and Technology (No.~2021ZD0301500).

\appendix
\section{Solution of dynamic evolution}
\label{App:solution}

In this section, we derive an analytic solution for dynamic evolution of the boson modes.

\subsection{One boson mode coupled to a channel}
We consider a system of one boson mode $a$ coupled to a channel $c$ with coupling parameter $g$. The Hamiltonian can be expressed in the interaction frame as
\begin{equation}
\label{simple Hamiltonian}
	\tilde{H}_{\rm int}=g \left(ac^\dagger+ca^\dagger+ace^{-2i\omega t}+a^\dagger c^\dagger e^{2i\omega t} \right).
\end{equation}
Here we retain the counterrotation terms $ace^{-2i\omega t}$ and $a^\dagger c^\dagger e^{2i\omega t}$ which are neglected if the rotating wave approximation (RWA) is applied. The modes $a$ and $c$ are assumed to be resonant with frequency $\omega$. The Heisenberg equation of the modes can be obtained as
\begin{eqnarray}
		\label{Heisenberg equation}
			\dot{a}&=&-ig \left(c+c^\dagger e^{2i\omega t}\right),\nonumber\\
			\dot{c}&=&-ig \left(a+a^\dagger e^{2i\omega t}\right).
\end{eqnarray}
By making the transformation $a(t)=A(t)e^{i\omega t}$ and $c(t)=C(t)e^{i\omega t}$, which is equivalent to transform from the interaction frame to the Schr\"{o}dinger frame, the equations of motion can be expressed as follows,
\begin{eqnarray}
		\dot{A}+i\omega A&=&-ig \left(C+C^\dagger\right),\; \dot{A}^\dagger-i\omega A^\dagger=ig \left(C+C^\dagger\right),\nonumber\\
		\dot{C}+i\omega C&=&-ig\left(A+A^\dagger\right),\; \dot{C}^\dagger-i\omega C^\dagger=ig \left(A+A^\dagger\right).
\end{eqnarray}
Then we can define position and momentum operators by the corresponding annihilation and creation operators,
\begin{eqnarray}
		X=A+A^\dagger,\;&& X_c=C+C^\dagger,\nonumber\\
		P=\left(A-A^\dagger\right)/i,\;&& P_c=\left(C-C^\dagger\right)/i.
\end{eqnarray}
With that, the equations of motion can be rewritten as the following ordinary differential equations (ODEs),
\begin{eqnarray}
	\label{ODE XP}
		\dot{X}=\omega P,\; &&\dot{P}+\omega X=-2gX_c,\nonumber\\
		\dot{X_c}=\omega P_c,\; &&\dot{P_c}+\omega X_c=-2gX.
\end{eqnarray}

This set of first-order ODEs can be solved via a general approach as described below. The vector $\overrightarrow{Y}(t)=(y_1(t),y_2(t),\cdots,y_n(t))^{\rm T}$ as the function of time $t$ is driven by the coefficient matrix $K$, i.e., $\dot{\overrightarrow{Y}}=K\overrightarrow{Y}$. Diagonalizing the matrix $K=S^{-1}\Lambda S$, we can reorganize the modes into the eigenmodes $\overrightarrow{Y'}$ with independent oscillation frequencies $\dot{\overrightarrow{Y'}}=\Lambda\overrightarrow{Y'}$. Thus, the solution of the reorganized modes can be obtained directly as $\overrightarrow{Y'}(t)=e^{\Lambda t}\overrightarrow{Y'}(0)$. The solution of the original modes can then be expressed as $\overrightarrow{Y}(t)=S^{-1}e^{\Lambda t}S\overrightarrow{Y}(0)$. In particular, the dynamic solution of the $i$-th mode can be achieved as follows (Einstein convention is applied here),
\begin{equation}
    y_i(t)=S^{-1}_{ij}e^{\Lambda t}_{jk}S_{km}y_m(0)=e^{\lambda_j t}({\rm diag}(S^{-1}_i))_{jk}S_{km}y_m(0).
\end{equation}
Here we rearrange the $i$-th row of the transformation matrix $S^{-1}$ into the diagonal form ${\rm diag}(S^{-1}_i)$, such that it can interchange position with the oscillation matrix $e^{\Lambda t}$.

Using the general procedure outlined above, the four eigenvalues of the ODEs~(\ref{ODE XP}) are $\pm i\sqrt{\omega^2-2g\omega}$ and $\pm i\sqrt{\omega^2+2g\omega}$. According to the initial conditions, we get the exact solution of the dynamic evolution of mode $a$,
\begin{eqnarray}
\label{solution}
	a(t)&=&\left(
		e^{i\sqrt{\omega^2-2g\omega}t},\ e^{-i\sqrt{\omega^2-2g\omega}t},\ e^{i\sqrt{\omega^2+2g\omega}t},\ e^{-i\sqrt{\omega^2+2g\omega}t}
	\right) M
\left(\begin{array}{c}
	a(0)\\a^\dagger(0)\\c(0)\\c^\dagger(0)
\end{array}\right) e^{i\omega t},\nonumber\\
M&=&\frac{1}{4}\left(\begin{array}{cccc}
	1-\frac{\zeta-1}{\sqrt{\zeta^2-2\zeta}}&\frac{1}{\sqrt{\zeta^2-2\zeta}}&-1+\frac{\zeta-1}{\sqrt{\zeta^2-2\zeta}}&-\frac{1}{\sqrt{\zeta^2-2\zeta}}\\
	1+\frac{\zeta-1}{\sqrt{\zeta^2-2\zeta}}&-\frac{1}{\sqrt{\zeta^2-2\zeta}}&-1-\frac{\zeta-1}{\sqrt{\zeta^2-2\zeta}}&\frac{1}{\sqrt{\zeta^2-2\zeta}}\\
	1-\frac{\zeta+1}{\sqrt{\zeta^2+2\zeta}}&-\frac{1}{\sqrt{\zeta^2+2\zeta}}&1-\frac{\zeta+1}{\sqrt{\zeta^2+2\zeta}}&-\frac{1}{\sqrt{\zeta^2+2\zeta}}\\
	1+\frac{\zeta+1}{\sqrt{\zeta^2+2\zeta}}&\frac{1}{\sqrt{\zeta^2+2\zeta}}&1+\frac{\zeta+1}{\sqrt{\zeta^2+2\zeta}}&\frac{1}{\sqrt{\zeta^2+2\zeta}}
\end{array}\right),
\end{eqnarray}
where $\zeta={\omega}/{g}$.
The solution for mode $c$ can be obtained analogously, just by exchanging $a$ and $c$ in the solution of mode $a$. Notice that in the case of $|g|>\omega/2$, we define $\sqrt{\omega^2-2g\omega}=i\sqrt{2g\omega-\omega^2}$ and $\sqrt{\zeta^2-2\zeta}=i\sqrt{2\zeta-\zeta^2}$ to avoid possible confusion of double-valued square root.

Some special attention needs to be paid for the case of $g=\pm\omega/2$, where the denominators of several elements in the matrix approach zero. To overcome this technical difficulty, we remind that the mapping from coupling strength $g$ to the dynamic evolution function should be continuous and analytic. Thus, we can take limitation to calculate the expression Eq.~(\ref{solution}) for $g\rightarrow\pm\omega/2$. For example, the result for the special point $g\rightarrow\omega/2$ reads
\begin{eqnarray}
	a(t) &=&\bigg\{\frac{a(0)}{2} \left[1+\cos(\sqrt{2}\omega t)-igt-i\frac{3}{2}\sin(\sqrt{2}\omega t)\right]\nonumber\\
	&&+\frac{c(0)}{2} \left[\cos(\sqrt{2}\omega t)-1+igt-i\frac{3}{2}\sin(\sqrt{2}\omega t)\right]\nonumber\\
	&&+\frac{a^\dagger(0)}{2} \left[igt-\frac{i}{2}\sin(\sqrt{2}\omega t)\right]\nonumber\\
	&&+\frac{c^\dagger(0)}{2} \left[-igt-\frac{i}{2}\sin(\sqrt{2}\omega t) \right]\bigg\}e^{i\omega t}.
\end{eqnarray}

For a small coupling strength $g$, or equivalently a large value of $\zeta$, the expression above can be rewritten as
\begin{eqnarray}
	a(t)&=&\bigg\{\frac{a(0)}{2} \bigg[\cos(\sqrt{\omega^2-2g\omega}t)+\cos(\sqrt{\omega^2+2g\omega}t)\nonumber\\
		&&-i\frac{\zeta-1}{\sqrt{\zeta^2-2\zeta}}\sin(\sqrt{\omega^2-2g\omega}t)-i\frac{\zeta+1}{\sqrt{\zeta^2+2\zeta}}\sin(\sqrt{\omega^2+2g\omega}t) \bigg]\nonumber\\
		&&+\frac{c(0)}{2}\bigg[\cos(\sqrt{\omega^2+2g\omega}t)-\cos(\sqrt{\omega^2-2g\omega}t)\\
		&&+i\frac{\zeta-1}{\sqrt{\zeta^2-2\zeta}}\sin(\sqrt{\omega^2-2g\omega}t)-i\frac{\zeta+1}{\sqrt{\zeta^2+2\zeta}}\sin(\sqrt{\omega^2+2g\omega}t)\bigg] \bigg\}e^{i\omega t}.\nonumber
\end{eqnarray}

Retain the first-order term of the frequency $\sqrt{\omega^2\pm 2g\omega}\approx \omega\pm g$, and set $gt={\pi}/{2}$, we can get the identical results with RWA,
\begin{eqnarray}
	a(t)&\approx&\frac{c(0)}{2} \left[\cos\left(\omega t+\frac{\pi}{2}\right)-\cos\left(\omega t-\frac{\pi}{2}\right)
	+i\sin\left(\omega t-\frac{\pi}{2} \right)-i\sin \left(\omega t+\frac{\pi}{2} \right)\right]e^{i\omega t}\nonumber\\
	&=&c(0) \left[\cos \left(\omega t+\frac{\pi}{2} \right)-i\sin\left(\omega t+\frac{\pi}{2}\right) \right]e^{i\omega t} \nonumber\\
	&=&c(0)e^{-i(\omega t+{\pi}/{2})}e^{i\omega t}\nonumber\\
	&=&-ic(0).
\end{eqnarray}

\subsection{Multiple modes coupled to an intermediate channel}
\label{App:multimode}

The more general case with multiple modes coupled to a single intermediate channel mode can be solved analogously. The Hamiltonian reads
\begin{equation}
	\tilde{H}_{\rm int}=\sum_{j=1}^{n}gk_j \left(a_jc^\dagger+ca_j^\dagger+a_jce^{-2i\omega t}+a_j^\dagger c^\dagger e^{2i\omega t} \right).
\end{equation}
By defining the united mode $a(t) \equiv \frac{\sum_{j=1}^n k_j a_j(t)}{\sqrt{\sum_{j=1}^n k_j^2}}$, and rearranging its distribution over individual boson modes, one can obtain the solutions of dynamic evolution as
\begin{eqnarray}
	a_i(t)&=& a_i(0)-\frac{k_i}{\sum_{j=1}^{n}k_j^2}\sum_{j=1}^{n}k_ja_j(0) +\frac{k_i}{\sum_{j=1}^{n}k_j^2}e^{i\omega t}\mu M \left(\begin{array}{c}
		\sum_{j=1}^{n}k_ja_j(0)\\\sum_{j=1}^{n}k_ja_j^\dagger(0)\\\sqrt{\sum_{j=1}^{n}k_j^2}c(0)\\\sqrt{\sum_{j=1}^{n}k_j^2}c^\dagger(0)
	\end{array}\right),\nonumber\\
	c(t)&=&\mu M
	\left(\begin{array}{c}
		c(0)\\c^\dagger(0)\\ \frac{\sum_{j=1}^{n}k_ja_j(0)}{\sqrt{\sum_{j=1}^{n}k_j^2}}\\\frac{\sum_{j=1}^{n}k_ja_j^\dagger(0)}{\sqrt{\sum_{j=1}^{n}k_j^2}}
	\end{array}\right)e^{i\omega t},\nonumber\\
        \mu&=&\left(\begin{array}{cccc}
		e^{i\sqrt{\omega^2-2g'\omega}t},&e^{-i\sqrt{\omega^2-2g'\omega}t},&e^{i\sqrt{\omega^2+2g'\omega}t},&e^{-i\sqrt{\omega^2+2g'\omega}t}
	\end{array}\right),\nonumber\\
	M&=&\frac{1}{4}\left(\begin{array}{cccc}
		1-\frac{\zeta-1}{\sqrt{\zeta^2-2\zeta}}&\frac{1}{\sqrt{\zeta^2-2\zeta}}&-1+\frac{\zeta-1}{\sqrt{\zeta^2-2\zeta}}&-\frac{1}{\sqrt{\zeta^2-2\zeta}}\\
		1+\frac{\zeta-1}{\sqrt{\zeta^2-2\zeta}}&-\frac{1}{\sqrt{\zeta^2-2\zeta}}&-1-\frac{\zeta-1}{\sqrt{\zeta^2-2\zeta}}&\frac{1}{\sqrt{\zeta^2-2\zeta}}\\
		1-\frac{\zeta+1}{\sqrt{\zeta^2+2\zeta}}&-\frac{1}{\sqrt{\zeta^2+2\zeta}}&1-\frac{\zeta+1}{\sqrt{\zeta^2+2\zeta}}&-\frac{1}{\sqrt{\zeta^2+2\zeta}}\\
		1+\frac{\zeta+1}{\sqrt{\zeta^2+2\zeta}}&\frac{1}{\sqrt{\zeta^2+2\zeta}}&1+\frac{\zeta+1}{\sqrt{\zeta^2+2\zeta}}&\frac{1}{\sqrt{\zeta^2+2\zeta}}
	\end{array}\right),
\end{eqnarray}
where $\zeta={\omega}/{g'}$ and $g'=\sqrt{\sum_{j=1}^{n}k_j^2}g$. Therefore, one can predict arbitrary correlation functions exactly.

\section{Optimized scheme of quantum state transfer}
\label{App:QST}

The conventional quantum state transfer (QST) scenario obtained under RWA suggests the pulse condition $g'\tau=\pi$, under which the fidelity of the task drops significantly in the strong-coupling regime. The exact solution obtained in the previous section can provide an optimized condition with better performance.

For the case of two bosonic nodes coupled to an intermediate channel with the same coupling strength, the exact solution of mode $a_1$ is expressed without any approximation as,
\begin{eqnarray}
    \label{a1 solution}
	a_1(t)&=&\Bigg\{ \frac{a_1(0)+a_2(0)}{4} \bigg[\cos \left(\sqrt{\omega^2-2g'\omega}t \right)+\cos \left(\sqrt{\omega^2+2g'\omega}t \right)\nonumber\\
	&& \hspace{1cm} -i\frac{\zeta-1}{\sqrt{\zeta^2-2\zeta}}\sin \left(\sqrt{\omega^2-2g'\omega}t \right)-i\frac{\zeta+1}{\sqrt{\zeta^2+2\zeta}}\sin \left(\sqrt{\omega^2+2g'\omega}t \right) \bigg]\nonumber\\
	&&+\frac{c(0)}{2\sqrt{2}} \bigg[\cos \left(\sqrt{\omega^2+2g'\omega}t \right)-\cos \left(\sqrt{\omega^2-2g'\omega}t \right)\nonumber\\
	&& \hspace{1cm} +i\frac{\zeta-1}{\sqrt{\zeta^2-2\zeta}}\sin \left(\sqrt{\omega^2-2g'\omega}t \right)-i\frac{\zeta+1}{\sqrt{\zeta^2+2\zeta}}\sin \left(\sqrt{\omega^2+2g'\omega}t \right) \bigg]\nonumber\\
	&&+\frac{a_1^\dagger(0)+a_2^\dagger(0)}{4} \left[\frac{i}{\sqrt{\zeta^2-2\zeta}}\sin(\sqrt{\omega^2-2g'\omega}t)-\frac{i}{\sqrt{\zeta^2+2\zeta}}\sin \left(\sqrt{\omega^2+2g'\omega}t \right) \right]\nonumber\\
	&&+\frac{c^\dagger(0)}{2\sqrt{2}} \left[-\frac{i}{\sqrt{\zeta^2-2\zeta}}\sin \left(\sqrt{\omega^2-2g'\omega}t \right)-\frac{i}{\sqrt{\zeta^2+2\zeta}}\sin \left(\sqrt{\omega^2+2g'\omega}t \right) \right] \Bigg\}e^{i\omega t} \nonumber\\
	&&+\frac{a_1(0)-a_2(0)}{2}\nonumber\\
	\label{K_11 def} &\equiv& K_{11}a_1(0)+K_{21}a_2(0)+K_{c1}c(0)+K_{12}a_1^\dagger(0)+K_{22}a_2^\dagger(0)+K_{c2}c^\dagger(0).
\end{eqnarray}

In the weak-coupling limit where RWA is valid ($\zeta\rightarrow+\infty$, $\sqrt{\omega^2\pm2g'\omega}\approx \omega\pm g'$), the solution predicts the final mode $a_1(\tau)=-a_2(0)$, i.e., the QST is ideal. However, in the strong-coupling regime where RWA breaks down, the final mode $a_1(\tau)$ deviates from $a_2(0)$ and causes an accumulation of infidelity. According to Eq.~(\ref{a1 solution}), we find the optimized pulse ansatz to achieve better QST performance as follows,
\begin{eqnarray}
\label{optimized pulse ansatz}
    &&\left(\sqrt{1+\frac{2}{\zeta}}-\sqrt{1-\frac{2}{\zeta}}\right) \theta=2\pi,\nonumber\\
	&&\sqrt{1+\frac{2}{\zeta}}\theta=m\pi,
\end{eqnarray}
where $\theta=\omega \tau$ is a dimensionless parameter dependent on the system frequency $\omega$ and the QST time $\tau$. The parameter $\zeta=\omega/g'$ determines the relative coupling strength. The pulse parameter $m$ is an integer which needs to be chosen appropriately in experiments. Under the optimized pulse condition above, we can easily check that the final mode can be simplified as $a_1(\tau)=K_{11}a_1(0)+K_{21}a_2(0)$, with $K_{c1}(\tau)=K_{12}(\tau)=K_{22}(\tau)=K_{c2}(\tau)=0$. The parameters $\zeta$ and $\theta$ can be solved from Eq.~(\ref{optimized pulse ansatz}) as
\begin{eqnarray}
\label{optimized pulse ansatz 2}
    \zeta&=&2\frac{(1-2/m)^2+1}{1-(1-2/m)^2},\nonumber\\
	\theta&=&m\pi\sqrt{\frac{(1-2/m)^2+1}{2}}.
\end{eqnarray}
This solution gives the optimized pulse scheme of QST. It is easy to find that a larger value of $m$ corresponds to a weaker coupling strength $g'$ and a longer QST time $\tau$. Specifically, a strong-coupling regime with $g'\gtrsim 0.1\omega$ is reached for $m\lesssim 11$. 

To illustrate the model more explicitly, we write down the potential energy part of the Hamiltonian with canonical variables as $V=\frac{1}{2}\omega(X_1^2+X_2^2+X_c^2)+2gX_c(X_1+X_2)$. This potential can be expressed in a diagonal form using eigenvectors $\{Y_1,Y_2,Y_3\}$, leading to $V=\sum_i\tilde{\omega}_iY_i^2$ with $\tilde{\omega}_{1,2,3}=\omega,\omega\pm2g'$. By taking the kinetic energy part $T=\frac{\omega}{2}(P_1^2+P_2^2+P_c^2)$ into account, the Hamiltonian can be diagonalized with the eigenmodes as $H=\sum_i\sqrt{\omega\tilde{\omega}_i}b_i^\dagger b_i$, whose eigenfrequencies $\sqrt{\omega\tilde{\omega}_i}$ are embodied in the dynamic solution~(\ref{a1 solution}). In the regime of $2\left|g'\right|<\omega$, i.e., $m>2$, the potential trap forms a three-dimensional parabolic surface opening up in all three directions, which establishes a bounded system with positive eigenfrequencies  $\sqrt{\omega\tilde{\omega}_i}$. While entering into the regime of $2\left|g'\right|\le\omega$, i.e., $m=1$ or $2$, one dimension of the parabolic surface turns to curve down. Such unbounded potential will cause unphysical imaginary eigenfrequency, which we should avoid.

\section{Conservation of the total excitations}
\label{App:conservation}

\begin{figure}[tb]
	\centering
	\includegraphics[width=\textwidth]{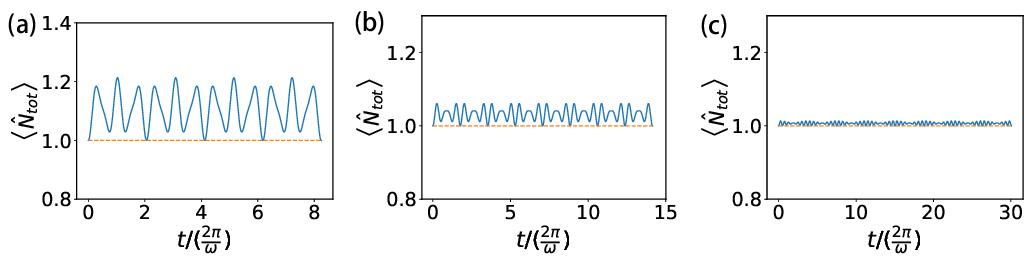}
	\caption{The total number of excitations $\left<\hat{N}_{\rm tot}\right>$ changes against time $t$ for different selections of $m$, (a) $m=5$; (b) $m=8$; (c) $m=16$. Here the blue solid (orange dashed) lines stands for the Hamiltonian with (without) the counterrotation terms. The initial excitations number is set as $\left<\hat{N}_{\rm tot}\right>=1$ in all cases.}
	\label{preserve excitations}
\end{figure}

We notice that using the optimized scheme of pulse, the operators $\left\{a_1(\tau),a_2(\tau),c(\tau)\right\}$ can be mapped into the subspace spanned only by the initial annihilation operators $\left\{a_1(0),a_2(0),c(0)\right\}$ (without the creation operators). Thus, we can define a transfer matrix $S$ to represent the map from the initial modes to the final modes $a_i(\tau)=\sum_jS_{ij}a_j(0)$. Apparently, the $S$ matrix is unitary according to the commutation relations $[{\cal O}_i(0), {\cal O}_j^\dagger(0)]=\delta_{ij}$ and $[{\cal O}_i(\tau), {\cal O}_j^\dagger(\tau)]=[U(\tau){\cal O}_i(0)U^\dagger(\tau),U(\tau){\cal O}_j^\dagger(0)U^\dagger(\tau)]=\delta_{ij}$, where ${\cal O}_{i=1,2,3} \in \{a_1, c, a_2\}$. Then, we can easily check the conservation of total excitations by calculating the expectation value $\left<N_{tot}\right>\equiv\left<\sum_i{\cal O}_i^\dagger{\cal O}_i\right>$.
\begin{eqnarray}
	\left<\sum_i{\cal O}_i^\dagger(\tau){\cal O}_i(\tau)\right>&=&\left<\sum_{i,k,l}S_{ik}^*{\cal O}_k^\dagger(0) S_{il}{\cal O}_l(0)\right>\nonumber\\
	&=&\left<\sum_{k,l}\delta_{kl}{\cal O}_k^\dagger(0){\cal O}_l(0)\right>\nonumber\\
	&=&\left<\sum_l{\cal O}_l^\dagger(0){\cal O}_l(0)\right>.
\end{eqnarray}
Therefore, the carefully selected pulse ansatz makes the creation and annihilation operator effectively decoupled, and the total excitation number can be preserved regardless of the breakdown of the U$(1)$ symmetry.

The change of the total excitations for cases of $m=5$, 8, and 16 are demonstrated in Fig.~\ref{preserve excitations}(a), \ref{preserve excitations}(b) and \ref{preserve excitations}(c), respectively, where we define $\hat{N}_{tot}\equiv a^\dagger_1a_1+a^\dagger_2a_2+c^\dagger c$. The blue solid (orange dashed) line represents the dynamics from the Hamiltonian with (without) the counterrotation terms. As the coupling strength is weak enough, such as the $m=16$ case, the orange dashed line deviates only very little from the blue solid line, indicating that the RWA can be safely applied. However, the total number of excitations manifest greater and non-negligible fluctuations in the strong-coupling regime for $m=5$. In another aspect, the fluctuation exhibits obvious periodicity. And the number of excitations keeps unchanged after a complete cycle. We set our QST condition right after one cycle to suppress the error brought by the fluctuation of excitations.

\section{Error estimation}
\label{App:QSL}

By implementing the solution of dynamic evolution to a practical QST task, we are able to give a prediction of the tradeoff relation between the task duration and error. 

We first consider a task of transferring a Fock state $\ket{n_1}$ from mode $a_1$ to $a_2$ via an ideal channel at zero temperature. When the QST process is completed at time $\tau$, the received state can be expressed as,
\begin{eqnarray}
\label{quantum speed limit}
	\ket{\Psi(\tau)}&=&U(\tau)\ket{n_1,0_c, 0_2} = U(\tau)\frac{(a_1^\dagger)^{n_1}}{\sqrt{n_1!}}\ket{0_1,0_c,0_2} \nonumber\\
	&=& \frac{(U(\tau)a_1^\dagger U^\dagger(\tau))^{n_1}}{\sqrt{n_1!}}U(\tau)\ket{0_1,0_c,0_2}\nonumber\\
	&=& \frac{(K_{11}a_1^\dagger+K_{21}a_2^\dagger)^{n_1}}{\sqrt{n_1!}}\ket{0_1,0_c,0_2}\nonumber\\	&=&\sum_{i=0}^{n_1}C_{n_1}^iK_{11}^iK_{21}^{n_1-i}\sqrt{\frac{i!(n_1-1)!}{n_1!}}\ket{i_1,0_c,(n_1-i)_2}\\
	&=&K_{21}^{n_1}\ket{0_1,0_c,(n_1)_2}+n_1\frac{1}{\sqrt{n_1}}K_{11}K_{21}^{n_1-1}\ket{1_1,0_c,(n_1-1)_2}+O(K_{11}^2K_{21}^{n_1-2}).\nonumber
\end{eqnarray}
The evolution operator is represented as $U(\tau)=\mathcal{T} {\rm exp}[-i\int_{0}^{\tau}dt\tilde{H}_{\rm int}(t)]$, where $\mathcal{T}$ stands for the time-ordered operator and the Hamiltonian is given by Eq.~(\ref{Hamiltonian_total}). 

Using the notation defined in the previous section, we can write down two identities 
\begin{eqnarray}
	U(\tau)\sum_{j,l}S^\dagger_{ij}{\cal O}_j(0)S^T_{il}{\cal O}_l^\dagger(0)U^\dagger(\tau) &=& \sum_{j,l}S^\dagger_{ij}{\cal O}_j(\tau)S^T_{il}{\cal O}_l^\dagger(\tau) \nonumber\\
	&=& \sum_{jk,lm}S^\dagger_{ij}S_{jk}{\cal O}_k(0)S^T_{il}S_{lm}^*{\cal O}_m^\dagger(0)
	\nonumber\\
	&=&\sum_{k,m}\delta_{ik}{\cal O}_k(0)\delta_{im}{\cal O}_m^\dagger(0)\nonumber\\
	&=&{\cal O}_i(0){\cal O}_i^\dagger(0), \label{iden1} \\
	\sum_{j,l}S^\dagger_{ij}{\cal O}_j(0)S^T_{il}{\cal O}_l^\dagger(0)\ket{0_1,0_c,0_2}&=&\sum_{j,l}S^\dagger_{ij}S^T_{il}\delta_{jl}\ket{0_1,0_c,0_2}\nonumber\\
	&=&\sum_{j}S^\dagger_{ij}S_{ji}\ket{0_1,0_c,0_2}\nonumber\\
	&=&\ket{0_1,0_c,0_2}, \label{iden2}
\end{eqnarray}
where $S^T$ is the transpose of $S$. Thus, the final state $U(\tau)\ket{0_1,0_c,0_2}$ can be obtained as
\begin{eqnarray}
	U(\tau)\ket{0_1,0_c,0_2}&=&U(\tau)\sum_{j,l}S^\dagger_{ij}{\cal O}_j(0)S^T_{il}{\cal O}_l^\dagger(0)\ket{0_1,0_c,0_2}\nonumber\\
	&=&U(\tau)\sum_{j,l}S^\dagger_{ij}{\cal O}_j(0)S^T_{il}{\cal O}_l^\dagger(0)U^\dagger(\tau)U(\tau)\ket{0_1,0_c,0_2}\nonumber\\
	&=&{\cal O}_i(0){\cal O}_i^\dagger(0)U(\tau)\ket{0_1,0_c,0_2}\nonumber\\
	&=&[{\cal O}_i^\dagger(0){\cal O}_i(0)+1]U(\tau)\ket{0_1,0_c,0_2}.
\end{eqnarray}
This expression then leads to another relation $0={\cal O}_i^\dagger(0){\cal O}_i(0)U(\tau)\ket{0_1,0_c,0_2}$. By expanding the state $U(\tau)\ket{0_1,0_c,0_2}$ into Fock basis as $U(\tau) \ket{0_1,0_c,0_2} = \sum_{n_i} A_{n_i} \ket{n_i}\otimes \ket{\psi_{n_i}}$, we get 
\begin{eqnarray}
0=\sum_{n_i} n_i A_{n_i} \ket{n_i} \otimes \ket{\psi_{n_i}}.
\end{eqnarray}
Here, $\ket{n_i}$ denotes the Fock state of mode ${\cal O}_i$, $\ket{\psi_{n_i}}$ is the state in the orthogonal subspace spanned by ${\cal O}_{j\neq i}$, and $A_{n_i}$ is the corresponding coefficient. Owing to the orthogonality of the Fock basis, we conclude that $A_{n_i \neq 0} = 0$ for all three modes ${\cal O}_i$, and thus $U(\tau)\ket{0_1,0_c,0_2}=\ket{0_1,0_c,0_2}$.

\begin{figure}[tb]
	\centering
	\includegraphics[width=0.8\textwidth]{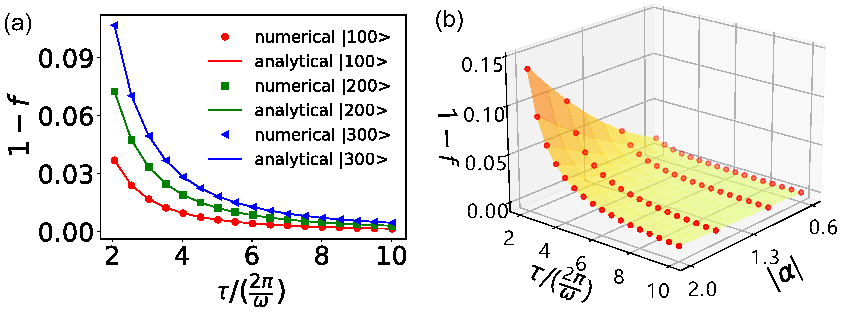}
	\caption{The tradeoff relation between the evolution time $\tau$ and the infidelity $1-f$. (a) Transferring Fock states $\ket{1}$ (red), $\ket{2}$ (green), $\ket{3}$ (blue). The solid lines represent the analytical results of the error, while the dots show numerical results of dynamic evolution using our optimized QST method. (b) Transferring coherent states with amplitude $\left|\alpha\right|=0.6\sim2.0$. The curved surface shows the analytical relation, while the dots around the surface are the numerical results. The range of $m$ is chosen from $5\sim17$.}
	\label{QSL}
\end{figure}

In the last line of Eq.~(\ref{quantum speed limit}), we expand the expression of the final state in orders of ${K_{11}}/{K_{21}}$, since the conditions of $K_{21}\sim 1$ and $K_{11}\sim 0$ are naturally expected to obtain high fidelity. To the first order, the error is
\begin{equation}
    1-f=\left|n_1\frac{1}{\sqrt{n_1}}K_{11}K_{21}^{n_1-1}\right|^2\approx\left|\sqrt{n_1}K_{11}\right|^2=n_1\left|K_{11}\right|^2.
\end{equation}
A straightforward estimation of the error in general cases gives $1-f=\left<n_1\right>\left|K_{11}\right|^2$. Using the definition Eq.~(\ref{K_11 def}), we obtain the exact result of $K_{11}$ as
\begin{eqnarray}
\label{expression of K11}
    K_{11}&=&\frac{1}{2}+\frac{1}{4} \bigg[ \cos\left(\sqrt{\omega^2-2g'\omega}t\right)+\cos\left(\sqrt{\omega^2+2g'\omega}t\right)\\
    && -i\frac{\zeta-1}{\sqrt{\zeta^2-2\zeta}}\sin\left(\sqrt{\omega^2-2g'\omega}t\right)-i\frac{\zeta+1}{\sqrt{\zeta^2+2\zeta}}\sin\left(\sqrt{\omega^2+2g'\omega}t\right) \bigg]e^{i\omega t}.\nonumber
\end{eqnarray}

In Eq.~(\ref{function relation}), the tradeoff relation is expressed as the constraint function $F(\rho,1-f,\tau)=0$. Obviously, the function can be acquired according to the analysis
\begin{equation}
    \label{function relation2}
    F(\rho,1-f,\tau)=1-f-\left<n_1\right>_\rho\left|K_{11}(t=\tau)\right|^2.
\end{equation}
According to the calculation above, the coefficient $\left|K_{11}\right|=\sin[-(\sqrt{1+{2}/{\zeta}}+\sqrt{1-{2}/{\zeta}}-2)\theta/4]$. Thus, we can define a function $G(m)\equiv \sin[-(\sqrt{1+{2}/{\zeta(m)}}+\sqrt{1-{2}/{\zeta(m)}}-2)\theta(m)/4]$, where $m$ is viewed as a continuous parameter temporarily. One can easily check that $G(m)$ is a monotonically decreasing function of $m$. 

Once the error that the receiver can tolerate is set as $E_{\rm tol}$, the QST time $\tau$ can be evaluated by Eq.~(\ref{function relation2}),
\begin{equation}
    1-f=\left<n_1\right>_\rho\left|K_{11}(t=\tau)\right|^2<E_{\rm tol}.
\end{equation}
Equivalently, it can be rewritten as $G(m)<\sqrt{{E_{\rm tol}}/{\left<n_1\right>_\rho}}$. By solving the inverse function of $G(m)$, we can get the threshold of parameter $m$, which reads $m>m_{\rm th}\equiv G^{-1}\left(\sqrt{{E_{\rm tol}}/{\left<n_1\right>_\rho}}\right)$. Thus, the threshold of $\tau$ can be obtained by $\theta_{\rm th}=\theta(\lfloor m_{\rm th}\rfloor+1)$, where $\lfloor m_{\rm th}\rfloor$ is the greatest integer less than or equal to $m_{\rm th}$. Since $\theta_{\rm th}\equiv\omega\tau_{\rm th}$, the shortest transfer time is given by $\tau_{\rm th}={\theta_{\rm th}}/{\omega}$.

For instance, the results for transferring a Fock state $\ket{n_1}$ with $n_1=1,2,3$ via a zero-temperature channel is displayed in Fig.~\ref{QSL}(a), while the ones for a coherent state $\ket{\alpha}$ with $\left|\alpha\right|=0.6\sim2.0$ is shown in Fig.~\ref{QSL}(b). Here the analytic results are calculated with the expression Eq.~(\ref{function relation2}), while the numerical results are obtained via a direct evolution of the equations of motion associated with the Hamiltonian without RWA. Notice that the analytic prediction agrees perfectly with the numerical results in all parameter regimes.

\begin{figure}[tb]
	\centering
	\includegraphics[width=0.8\textwidth]{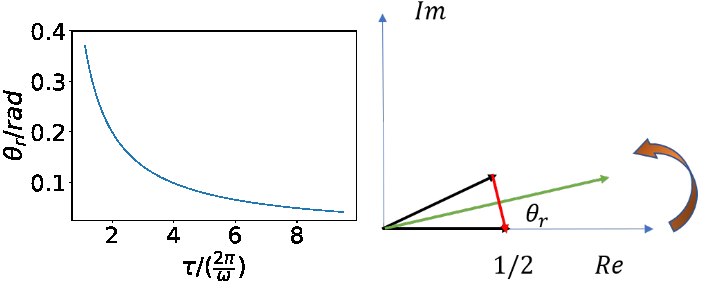}
	\caption{The left figure displays the rotation angle $\theta_r$ changes against the QST time $\tau$. The right figure illustrates the phase error originated from strong coupling. The addition of the two black vectors gives the coefficient $K_{21}$(green vector), while the difference gives the coefficient $K_{11}$ (red vector). As the parameter $m$ gets smaller, the upper black vector will rotate counterclockwise (the direction of the red thick arrow), making the green vector rotate counterclockwise and leading to a larger value of $\theta_r$.}
    \label{suppfigs}
\end{figure}

\section{The extra rotation in phase space}
\label{App:rotation}

Strong coupling effect induces a non-negligible rotation of the initial state in phase space. Using our optimized pulse ansatz Eq.~(\ref{optimized pulse ansatz 2}), we can obtain the final state,
\begin{eqnarray}
    \label{rotation}
	a_1(t)&=&\frac{a_1(0)+a_2(0)}{4}\bigg[\cos\left(\sqrt{\omega^2-2g'\omega}t\right)+\cos\left(\sqrt{\omega^2+2g'\omega}t\right)\nonumber\\
	&&-i\frac{\zeta-1}{\sqrt{\zeta^2-2\zeta}}\sin\left(\sqrt{\omega^2-2g'\omega}t\right)-i\frac{\zeta+1}{\sqrt{\zeta^2+2\zeta}}\sin\left(\sqrt{\omega^2+2g'\omega}t\right)\bigg]
	e^{i\omega t}\nonumber\\
	&&+\frac{a_1(0)-a_2(0)}{2}\nonumber\\
	&=&\frac{a_1(0)+a_2(0)}{2}\left[\cos\left(\sqrt{\omega^2-2g'\omega}t\right)-i\sin\left(\sqrt{\omega^2-2g'\omega}t\right)\right]e^{i\omega t}+\frac{a_1(0)-a_2(0)}{2}\nonumber\\
	&=&\frac{a_1(0)+a_2(0)}{2}e^{-i\sqrt{1-{2}/{\zeta}}\omega t}e^{i\omega t}+\frac{a_1(0)-a_2(0)}{2}\nonumber\\
	&=&\frac{a_1(0)+a_2(0)}{2}e^{-i[(\sqrt{1+{2}/{\zeta}}+\sqrt{1-{2}/{\zeta}})/2-(\sqrt{1+{2}/{\zeta}}-\sqrt{1-{2}/{\zeta}})/2]\omega t}e^{i\omega t}+\frac{a_1(0)-a_2(0)}{2}\nonumber\\
	&=&-\frac{a_1(0)+a_2(0)}{2}e^{-i[(\sqrt{1+{2}/{\zeta}}+\sqrt{1-{2}/{\zeta}})/2-1]\omega t}+\frac{a_1(0)-a_2(0)}{2}\nonumber\\
	&=&-\frac{1}{2}\left(e^{-i[(\sqrt{1+{2}/{\zeta}}+\sqrt{1-{2}/{\zeta}})/2-1]\omega t}+1\right)a_2(0)\nonumber\\
	&&+\frac{1}{2}\left(1-e^{-i[(\sqrt{1+{2}/{\zeta}}+\sqrt{1-{2}/{\zeta}})/2-1]\omega t}\right)a_1(0)\nonumber\\
	&=&-\cos(\theta_r)e^{i\theta_r}a_2(0)+\sin(\theta_r)e^{i(\theta_r-\frac{\pi}{2})}a_1(0)\nonumber\\
	&=&K_{21}a_2(0)+K_{11}a_1(0),
\end{eqnarray}
where $\theta_r=-(\sqrt{1+{2}/{\zeta}}+\sqrt{1-{2}/{\zeta}}-2)\theta/4$, and $\left|K_{11}\right|=\sin(\theta_r)$. The rotation in the first term of Eq.~(\ref{rotation}) is the intrinsic result associated with the energy shift from $\omega\pm g'$ in the weak-coupling regime to $\sqrt{\omega^2\pm2g'\omega}$ in the strong-coupling regime. The second term is the error deviated from the desired mode $a_2(0)$, and the coefficient is given exactly as in Eq.~(\ref{expression of K11}). 

We can calculate the coefficients above according to a geometry relation in the complex plane. As illustrated in Fig.~\ref{suppfigs}, the rotation angle $\theta_r$ increases with the decreasing of QST time $\theta=\omega \tau$. Such behavior can be understood by the red arrow in the right subfigure. The black vector lying on the real axis stands for the number ${1}/{2}$, and the other black vector above the real axis refers to $({1}/{2}) e^{-i[(\sqrt{1+{2}/{\zeta}}+\sqrt{1-{2}/{\zeta}})/2-1]\omega t}=({1}/{2})e^{2i\theta_r}$. The superposition of the two vectors gives the coefficient $\cos(\theta_r)e^{i\theta_r}$ corresponding to the green vector in the subfigure, and the difference of the two vectors (red vector) gives the coefficients $K_{11}$ corresponding to the error of the QST task. Apparently, as the parameter $m$ gets smaller, the QST time becomes smaller. As a result, the second black rector will rotate more counterclockwise, making the length of the red vector (error) become lager. Meanwhile, the green vector will rotate more counterclockwise, leading to a larger angular shift in phase space.

\begin{figure}[tb]
	\centering
	\includegraphics[width=0.5\textwidth]{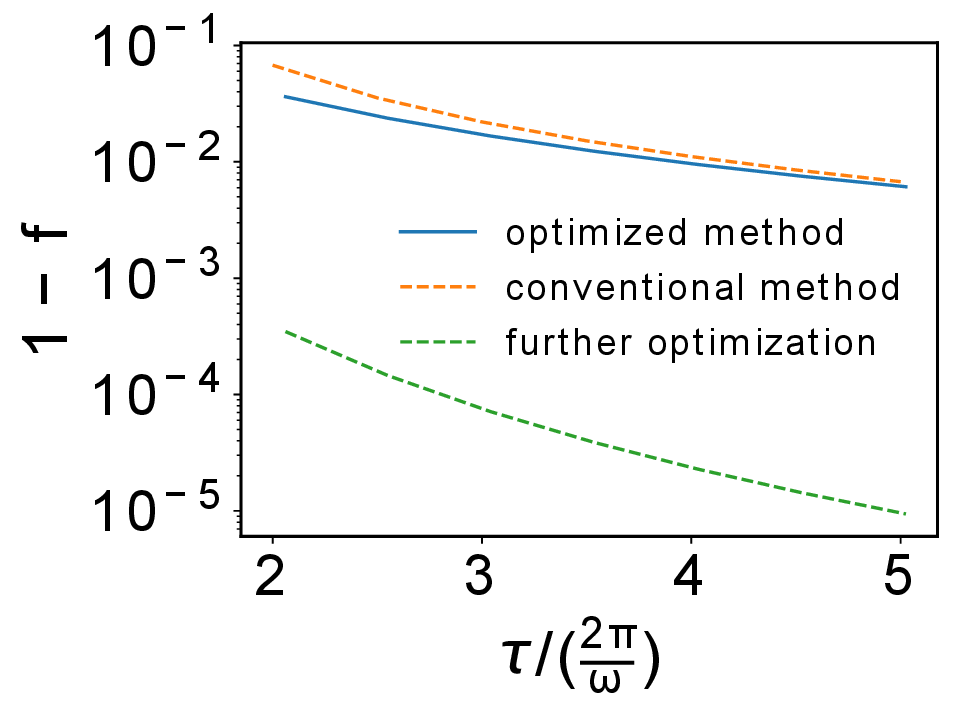}
	\caption{The infidelity when transferring a coherent state $\ket{\psi}=\ket{e^{i{\pi}/{2}}}$ through an ideal channel in the strong-coupling regime with $5\le m\le11$. It is found that the optimized method without further optimization (blue solid line) suppresses the infidelity in the conventional RWA method (orange dashed line), while a more remarkable decrease is observed after applying the further optimization with local rotation (green dashed line).}
	    \label{suppfig(f)}
\end{figure}

With that, an additional local rotation applied to node $a_2$ is proposed after the optimized QST pulse of Eq.~(\ref{optimized pulse ansatz 2}), in order to further suppress the infidelity caused by the rotation $\theta_r$. It can be realized by implementing a local pulse on node $a_2$ expressed as $H_r=g_r(t)a_2^\dagger a_2$ with the pulse area $\int g_r(t)dt=\theta_r$. 
Here, we take an example of transferring a coherent state $\ket{\psi}=\ket{e^{i{\pi}/{2}}}$ in the strong-coupling regime $g'\gtrsim 0.1\omega$ (corresponds to $5\le m\le11$ in our optimized QST method). As shown in Fig.~\ref{suppfig(f)}, such a local rotation, labeled as ``further optimization", can suppress the infidelity for $2\sim 3$ orders of magnitude, which is quite remarkable considering the error is already less than a few percent.

We emphasize that this further optimization method is efficient for coherent states since they are sensitive to the rotated phases in phase space. Thus, it is also useful to suppress the QST error when transferring a general state that is sensitive to phase, such as squeezed states, cat states, etc. On the other hand, this method makes no difference to Fock states, which are centrally symmetric in phase space and thus insensitive to a phase rotation.

\bibliography{ref}

\providecommand{\newblock}{}
\begin{thebibliography}{10}
\expandafter\ifx\csname url\endcsname\relax
  \def\url#1{{\tt #1}}\fi
\expandafter\ifx\csname urlprefix\endcsname\relax\def\urlprefix{URL }\fi
\providecommand{\eprint}[2][]{\url{#2}}

\bibitem{doi:10.1126/science.1231692}
Spring J~B, Metcalf B~J, Humphreys P~C, Kolthammer W~S, Jin X~M, Barbieri M,
  Datta A, Thomas-Peter N, Langford N~K, Kundys D {\em et~al.\/} 2013 {\em
  Science\/} {\bf 339} 798--801
  \urlprefix\url{https://www.science.org/doi/abs/10.1126/science.1231692}

\bibitem{RN94}
Tillmann M, Dakić B, Heilmann R, Nolte S, Szameit A and Walther P 2013 {\em
  Nat. Photon.\/} {\bf 7} 540--544
  \urlprefix\url{https://doi.org/10.1038/nphoton.2013.102}

\bibitem{RN95}
Spagnolo N, Vitelli C, Bentivegna M, Brod D~J, Crespi A, Flamini F, Giacomini
  S, Milani G, Ramponi R, Mataloni P {\em et~al.\/} 2014 {\em Nat. Photon.\/}
  {\bf 8} 615--620 \urlprefix\url{https://doi.org/10.1038/nphoton.2014.135}

\bibitem{PhysRevLett.123.250503}
Wang H, Qin J, Ding X, Chen M~C, Chen S, You X, He Y~M, Jiang X, You L, Wang Z
  {\em et~al.\/} 2019 {\em Phys. Rev. Lett.\/} {\bf 123} 250503
  \urlprefix\url{https://link.aps.org/doi/10.1103/PhysRevLett.123.250503}

\bibitem{RN96}
Arrazola J~M, Bergholm V, Brádler K, Bromley T~R, Collins M~J, Dhand I,
  Fumagalli A, Gerrits T, Goussev A, Helt L~G {\em et~al.\/} 2021 {\em
  Nature\/} {\bf 591} 54--60
  \urlprefix\url{https://doi.org/10.1038/s41586-021-03202-1}

\bibitem{PhysRevLett.119.170501}
Hamilton C~S, Kruse R, Sansoni L, Barkhofen S, Silberhorn C and Jex I 2017 {\em
  Phys. Rev. Lett.\/} {\bf 119} 170501
  \urlprefix\url{https://link.aps.org/doi/10.1103/PhysRevLett.119.170501}

\bibitem{RN93}
Knill E, Laflamme R and Milburn G~J 2001 {\em Nature\/} {\bf 409} 46--52
  \urlprefix\url{https://doi.org/10.1038/35051009}

\bibitem{RevModPhys.79.135}
Kok P, Munro W~J, Nemoto K, Ralph T~C, Dowling J~P and Milburn G~J 2007 {\em
  Rev. Mod. Phys.\/} {\bf 79} 135--174
  \urlprefix\url{https://link.aps.org/doi/10.1103/RevModPhys.79.135}

\bibitem{PhysRevLett.73.58}
Reck M, Zeilinger A, Bernstein H~J and Bertani P 1994 {\em Phys. Rev. Lett.\/}
  {\bf 73} 58--61
  \urlprefix\url{https://link.aps.org/doi/10.1103/PhysRevLett.73.58}

\bibitem{PhysRevLett.118.133601}
Vermersch B, Guimond P~O, Pichler H and Zoller P 2017 {\em Phys. Rev. Lett.\/}
  {\bf 118} 133601
  \urlprefix\url{https://link.aps.org/doi/10.1103/PhysRevLett.118.133601}

\bibitem{RN71}
Axline C~J, Burkhart L~D, Pfaff W, Zhang M, Chou K, Campagne-Ibarcq P, Reinhold
  P, Frunzio L, Girvin S~M, Jiang L, Devoret M~H and Schoelkopf R~J 2018 {\em
  Nat. Phys.\/} {\bf 14} 705--710
  \urlprefix\url{https://doi.org/10.1038/s41567-018-0115-y}

\bibitem{PRXQuantum.2.030321}
Burkhart L~D, Teoh J~D, Zhang Y, Axline C~J, Frunzio L, Devoret M, Jiang L,
  Girvin S and Schoelkopf R 2021 {\em PRX Quantum\/} {\bf 2} 030321
  \urlprefix\url{https://link.aps.org/doi/10.1103/PRXQuantum.2.030321}

\bibitem{PhysRevX.7.011035}
Xiang Z~L, Zhang M, Jiang L and Rabl P 2017 {\em Phys. Rev. X\/} {\bf 7} 011035
  \urlprefix\url{https://link.aps.org/doi/10.1103/PhysRevX.7.011035}

\bibitem{PhysRevApplied.17.054021}
Ai H, Fang Y~Y, Feng C~R, Peng Z and Xiang Z~L 2022 {\em Phys. Rev. Appl.\/}
  {\bf 17} 054021
  \urlprefix\url{https://link.aps.org/doi/10.1103/PhysRevApplied.17.054021}

\bibitem{PhysRevLett.130.050801}
Xiang Z~L, Olivares D~G, Garc\'{\i}a-Ripoll J~J and Rabl P 2023 {\em Phys. Rev.
  Lett.\/} {\bf 130} 050801
  \urlprefix\url{https://link.aps.org/doi/10.1103/PhysRevLett.130.050801}

\bibitem{PhysRevLett.108.153604}
Tian L 2012 {\em Phys. Rev. Lett.\/} {\bf 108} 153604
  \urlprefix\url{https://link.aps.org/doi/10.1103/PhysRevLett.108.153604}

\bibitem{PhysRevA.87.012339}
Qin W, Wang C and Long G~L 2013 {\em Phys. Rev. A\/} {\bf 87} 012339
  \urlprefix\url{https://link.aps.org/doi/10.1103/PhysRevA.87.012339}

\bibitem{Kimble}
Kimble H~J 2008 {\em Nature\/} {\bf 453} 1023--30
  \urlprefix\url{https://doi.org/10.1038/nature07127}

\bibitem{RN70}
Northup T~E and Blatt R 2014 {\em Nat. Photon.\/} {\bf 8} 356--363
  \urlprefix\url{https://doi.org/10.1038/nphoton.2014.53}

\bibitem{RevModPhys.82.1041}
Hammerer K, S\o{}rensen A~S and Polzik E~S 2010 {\em Rev. Mod. Phys.\/} {\bf
  82} 1041--1093
  \urlprefix\url{https://link.aps.org/doi/10.1103/RevModPhys.82.1041}

\bibitem{PhysRevLett.78.3221}
Cirac J~I, Zoller P, Kimble H~J and Mabuchi H 1997 {\em Phys. Rev. Lett.\/}
  {\bf 78} 3221--3224
  \urlprefix\url{https://link.aps.org/doi/10.1103/PhysRevLett.78.3221}

\bibitem{Ritter}
Ritter S, Nölleke C, Hahn C, Reiserer A, Neuzner A, Uphoff M, Mücke M,
  Figueroa E, Bochmann J and Rempe G 2012 {\em Nature\/} {\bf 484} 195--200
  \urlprefix\url{https://doi.org/10.1038/nature11023}

\bibitem{PhysRevA.50.R3589}
Zeng H and Lin F 1994 {\em Phys. Rev. A\/} {\bf 50} R3589--R3592
  \urlprefix\url{https://link.aps.org/doi/10.1103/PhysRevA.50.R3589}

\bibitem{PhysRevApplied.10.054009}
Li X, Ma Y, Han J, Chen T, Xu Y, Cai W, Wang H, Song Y, Xue Z~Y, Yin Z~q and
  Sun L 2018 {\em Phys. Rev. Appl.\/} {\bf 10} 054009
  \urlprefix\url{https://link.aps.org/doi/10.1103/PhysRevApplied.10.054009}

\bibitem{PhysRevB.84.014510}
Korotkov A~N 2011 {\em Phys. Rev. B\/} {\bf 84} 014510
  \urlprefix\url{https://link.aps.org/doi/10.1103/PhysRevB.84.014510}

\bibitem{PhysRevLett.95.170501}
Zhang J, Xie C and Peng K 2005 {\em Phys. Rev. Lett.\/} {\bf 95} 170501
  \urlprefix\url{https://link.aps.org/doi/10.1103/PhysRevLett.95.170501}

\bibitem{Xiao-Ling}
He X~L, Zheng Z~F, Zhang Y and Yang C~P 2020 {\em Quantum Inf. Process.\/} {\bf
  19} 80 \urlprefix\url{https://doi.org/10.1007/s11128-020-2578-x}

\bibitem{Brown}
Brown K~R, Ospelkaus C, Colombe Y, Wilson A~C, Leibfried D and Wineland D~J
  2011 {\em Nature\/} {\bf 471} 196--9
  \urlprefix\url{https://doi.org/10.1038/nature09721}

\bibitem{PhysRevA.92.053804}
Filip R and Rakhubovsky A~A 2015 {\em Phys. Rev. A\/} {\bf 92} 053804
  \urlprefix\url{https://link.aps.org/doi/10.1103/PhysRevA.92.053804}

\bibitem{Zeng:20}
Zeng Y~X, Shen J, Ding M~S and Li C 2020 {\em Opt. Express\/} {\bf 28}
  9587--9602
  \urlprefix\url{http://www.osapublishing.org/oe/abstract.cfm?URI=oe-28-7-9587}

\bibitem{PhysRevA.86.021801}
Singh S, Jing H, Wright E~M and Meystre P 2012 {\em Phys. Rev. A\/} {\bf 86}
  021801(R) \urlprefix\url{https://link.aps.org/doi/10.1103/PhysRevA.86.021801}

\bibitem{Parkins_1999}
Parkins A~S and Kimble H~J 1999 {\em J. Opt. B\/} {\bf 1} 496--504
  \urlprefix\url{https://doi.org/10.1088/1464-4266/1/4/323}

\bibitem{Barzanjeh2022Optomechanics}
Barzanjeh S, Xuereb A, Gr\"oblacher S, Paternostro M, Regal C~A and Weig E~M
  2022 {\em Nat. Phys.\/} {\bf 18} 15--24
  \urlprefix\url{https://doi.org/10.1038/s41567-021-01402-0}

\bibitem{Kane}
Kane B~E 1998 {\em Nature\/} {\bf 393} 133--137
  \urlprefix\url{https://doi.org/10.1038/30156}

\bibitem{PhysRevLett.106.040505}
Yao N~Y, Jiang L, Gorshkov A~V, Gong Z~X, Zhai A, Duan L~M and Lukin M~D 2011
  {\em Phys. Rev. Lett.\/} {\bf 106} 040505
  \urlprefix\url{https://link.aps.org/doi/10.1103/PhysRevLett.106.040505}

\bibitem{Kandel}
Kandel Y~P, Haifeng Q, Saeed F, Gardner G~C, Manfra M~J and Nichol J~M 2021
  {\em Nat. Commun.\/} {\bf 12} 2156
  \urlprefix\url{https://doi.org/10.1038/s41467-021-22416-5}

\bibitem{PhysRevLett.99.093901}
Wilson-Rae I, Nooshi N, Zwerger W and Kippenberg T~J 2007 {\em Phys. Rev.
  Lett.\/} {\bf 99} 093901
  \urlprefix\url{https://link.aps.org/doi/10.1103/PhysRevLett.99.093901}

\bibitem{RN97}
Neeley M, Bialczak R~C, Lenander M, Lucero E, Mariantoni M, O’Connell A~D,
  Sank D, Wang H, Weides M, Wenner J {\em et~al.\/} 2010 {\em Nature\/} {\bf
  467} 570--573 \urlprefix\url{https://doi.org/10.1038/nature09418}

\bibitem{RN98}
Ristè D, Dukalski M, Watson C~A, de~Lange G, Tiggelman M~J, Blanter Y~M,
  Lehnert K~W, Schouten R~N and DiCarlo L 2013 {\em Nature\/} {\bf 502}
  350--354 \urlprefix\url{https://doi.org/10.1038/nature12513}

\bibitem{RN100}
Yao X~C, Wang T~X, Xu P, Lu H, Pan G~S, Bao X~H, Peng C~Z, Lu C~Y, Chen Y~A and
  Pan J~W 2012 {\em Nat. Photon.\/} {\bf 6} 225--228
  \urlprefix\url{https://doi.org/10.1038/nphoton.2011.354}

\bibitem{RN101}
Jing B, Wang X~J, Yu Y, Sun P~F, Jiang Y, Yang S~J, Jiang W~H, Luo X~Y, Zhang
  J, Jiang X {\em et~al.\/} 2019 {\em Nat. Photon.\/} {\bf 13} 210--213
  \urlprefix\url{https://doi.org/10.1038/s41566-018-0342-x}

\bibitem{PhysRevA.50.R2799}
Cirac J~I and Zoller P 1994 {\em Phys. Rev. A\/} {\bf 50} R2799--R2802
  \urlprefix\url{https://link.aps.org/doi/10.1103/PhysRevA.50.R2799}

\bibitem{Ockeloen-Korppi}
Ockeloen-Korppi C, Damskägg E, Pirkkalainen J~M, Asjad M, Clerk A~A, Massel F,
  Woolley M~J and Sillanpää M 2018 {\em Nature\/} {\bf 556} 478--482
  \urlprefix\url{https://doi.org/10.1038/s41586-018-0038-x}

\bibitem{PhysRevA.84.052327}
Hofer S~G, Wieczorek W, Aspelmeyer M and Hammerer K 2011 {\em Phys. Rev. A\/}
  {\bf 84} 052327
  \urlprefix\url{https://link.aps.org/doi/10.1103/PhysRevA.84.052327}

\bibitem{RevModPhys.86.1391}
Aspelmeyer M, Kippenberg T~J and Marquardt F 2014 {\em Rev. Mod. Phys.\/} {\bf
  86} 1391--1452
  \urlprefix\url{https://link.aps.org/doi/10.1103/RevModPhys.86.1391}

\bibitem{LiOuLeiLiu+2021+2799+2832}
Li B~B, Ou L, Lei Y and Liu Y~C 2021 {\em Nanophotonics\/} {\bf 10} 2799--2832
  \urlprefix\url{https://doi.org/10.1515/nanoph-2021-0256}

\bibitem{RN82}
Gu X, Kockum A~F, Miranowicz A, Liu Y~x and Nori F 2017 {\em Phys. Rep.\/} {\bf
  718-719} 1--102
  \urlprefix\url{https://www.sciencedirect.com/science/article/pii/S0370157317303290}

\bibitem{jozsa1994fidelity}
Jozsa R 1994 {\em J. Mod. Opt.\/} {\bf 41} 2315
  \urlprefix\url{https://doi.org/10.1080/09500349414552171}

\bibitem{ourjoumtsev2006generating}
Ourjoumtsev A, Tualle-Brouri R, Laurat J and Grangier P 2006 {\em Science\/}
  {\bf 312} 83--86
  \urlprefix\url{https://www.science.org/doi/abs/10.1126/science.1122858}

\bibitem{PhysRevLett.101.233605}
Takahashi H, Wakui K, Suzuki S, Takeoka M, Hayasaka K, Furusawa A and Sasaki M
  2008 {\em Phys. Rev. Lett.\/} {\bf 101} 233605
  \urlprefix\url{https://link.aps.org/doi/10.1103/PhysRevLett.101.233605}

\bibitem{han2023remote}
Han D, Sun F, Wang N, Xiang Y, Wang M, Tian M, He Q and Su X {\em Laser
  Photonics Rev.\/}  2300103
  \urlprefix\url{https://onlinelibrary.wiley.com/doi/abs/10.1002/lpor.202300103}

\bibitem{PhysRevLett.116.163602}
Liao J~Q and Tian L 2016 {\em Phys. Rev. Lett.\/} {\bf 116} 163602
  \urlprefix\url{https://link.aps.org/doi/10.1103/PhysRevLett.116.163602}

\bibitem{Yang:18}
Yang C~P and Zheng Z~F 2018 {\em Opt. Lett.\/} {\bf 43} 5126--5129
  \urlprefix\url{http://opg.optica.org/ol/abstract.cfm?URI=ol-43-20-5126}

\bibitem{doi:10.1126/science.272.5265.1131}
Monroe C, Meekhof D~M, King B~E and Wineland D~J 1996 {\em Science\/} {\bf 272}
  1131--1136
  \urlprefix\url{https://www.science.org/doi/abs/10.1126/science.272.5265.1131}

\bibitem{doi:10.1126/science.aay0600}
Song C, Xu K, Li H, Zhang Y~R, Zhang X, Liu W, Guo Q, Wang Z, Ren W, Hao J {\em
  et~al.\/} 2019 {\em Science\/} {\bf 365} 574--577
  \urlprefix\url{https://www.science.org/doi/abs/10.1126/science.aay0600}

\bibitem{PhysRevLett.127.087203}
Sun F~X, Zheng S~S, Xiao Y, Gong Q, He Q and Xia K 2021 {\em Phys. Rev.
  Lett.\/} {\bf 127} 087203
  \urlprefix\url{https://link.aps.org/doi/10.1103/PhysRevLett.127.087203}

\bibitem{PhysRevLett.103.240501}
Caneva T, Murphy M, Calarco T, Fazio R, Montangero S, Giovannetti V and Santoro
  G~E 2009 {\em Phys. Rev. Lett.\/} {\bf 103} 240501
  \urlprefix\url{https://link.aps.org/doi/10.1103/PhysRevLett.103.240501}

\bibitem{PhysRevLett.118.150503}
Li J, Yang X, Peng X and Sun C~P 2017 {\em Phys. Rev. Lett.\/} {\bf 118}(15)
  150503
  \urlprefix\url{https://link.aps.org/doi/10.1103/PhysRevLett.118.150503}

\bibitem{RN90}
Gottesman D and Chuang I~L 1999 {\em Nature\/} {\bf 402} 390--393
  \urlprefix\url{https://doi.org/10.1038/46503}

\bibitem{RN91}
Chou K~S, Blumoff J~Z, Wang C~S, Reinhold P~C, Axline C~J, Gao Y~Y, Frunzio L,
  Devoret M~H, Jiang L and Schoelkopf R~J 2018 {\em Nature\/} {\bf 561}
  368--373 \urlprefix\url{https://doi.org/10.1038/s41586-018-0470-y}

\bibitem{doi:10.1126/science.aaw9415}
Wan Y, Kienzler D, Erickson S~D, Mayer K~H, Tan T~R, Wu J~J, Vasconcelos H~M,
  Glancy S, Knill E, Wineland D~J {\em et~al.\/} 2019 {\em Science\/} {\bf 364}
  875--878
  \urlprefix\url{https://www.science.org/doi/abs/10.1126/science.aaw9415}

\bibitem{PhysRevA.65.032314}
Vidal G and Werner R~F 2002 {\em Phys. Rev. A\/} {\bf 65} 032314
  \urlprefix\url{https://link.aps.org/doi/10.1103/PhysRevA.65.032314}

\end{thebibliography}
	
\end{document}